\documentclass[a4paper,twocolumn,11pt,accepted=2025-09-17]{quantumarticle}
\pdfoutput=1
\usepackage[utf8]{inputenc}
\usepackage[english]{babel}
\usepackage[T1]{fontenc}
\usepackage{amsmath}

\usepackage{tikz}
\usetikzlibrary{tikzmark,calc,patterns}
\usepackage{makecell}

\usepackage{mathtools}
\usepackage{hyperref}
\usepackage{ulem}
\usepackage{tikz}
\usepackage{lipsum}
\usepackage{amsmath}
\usepackage{amssymb}
\usepackage{fixltx2e}
\usepackage[numbers,sort&compress]{natbib}
\usepackage{graphicx} % Required for inserting images
\usepackage{physics}

\usepackage{xcolor}

\begin{document}
\title{Quantum Error Correction resilient against Atom Loss}
\author{Hugo Perrin}
\email{hugo.perrin@qperfect.io}
\affiliation{University of Strasbourg and CNRS, CESQ and ISIS (UMR 7006), aQCess, 67000 Strasbourg, France}
\affiliation{QPerfect SAS, 67200 Strasbourg, France}

\author{Sven Jandura}
\affiliation{University of Strasbourg and CNRS, CESQ and ISIS (UMR 7006), aQCess, 67000 Strasbourg, France}

\author{Guido Pupillo}
\affiliation{University of Strasbourg and CNRS, CESQ and ISIS (UMR 7006), aQCess, 67000 Strasbourg, France}
\affiliation{QPerfect SAS, 67200 Strasbourg, France}
\email{pupillo@unistra.fr}

\begin{abstract}
We investigate  quantum error correction protocols for neutral atoms quantum processors in the presence of atom loss. We complement the surface code with loss detection units (LDU) and analyze its performances by means of circuit-level simulations for two distinct protocols -- the standard LDU and a teleportation-based LDU --, focussing on the impact of both atom loss and depolarizing noise on the logical error probability. We introduce and employ a new adaptive decoding procedure that leverages the knowledge of loss locations provided by the LDUs, improving logical error probabilities by nearly three orders of magnitude compared to a naive decoder. For the considered error models, our results demonstrate the existence of an error threshold line that depends linearly on the probabilities of atom loss and of depolarizing errors. For zero depolarizing noise, the atom loss threshold is about $2.6\%$.  
\end{abstract}
\maketitle
\section{Introduction}
 Quantum error correction (QEC) is essential for realizing robust quantum memories and fault-tolerant quantum computing. Since its theoretical foundation in the 1990s~\cite{Shor1995,Gottesman1997,Knill1997,Preskill1998,Steane1999}, numerous protocols have been developed to achieve QEC for different, often rather abstract,  noise models. Recent experimental demonstrations of QEC in several quantum computing platforms~\cite{Egan2021,Ryan-anderson2021,Abobeih2022,Postler2022,Evered2023,Dasilva2024,Bluvstein2024,Finkelstein2024,Acharya2025} are now motivating %. As these experimental platforms are further developed, there is a pressing need to 
 the development of QEC strategies that are tailored to the specific noise sources of each individual platform. %, in order to accurately evaluate their future performance.

 Neutral atom quantum processors have recently emerged as a leading technology for large-scale fault tolerant quantum computing~\cite{Evered2023,Bluvstein2024,Finkelstein2024,Saffman2010,Browaeys2020,Henriet2020,Morgado2021,Graham2022,Anand2024,Scholl2023,Ma2023,Cao2024,Reichardt2024}. They are characterized by the possibility to encode qubits in internal states of the atoms with very long coherence times (up to tens of seconds), trap qubits in  arbitrary geometries using optical tweezer arrays, dynamical reconfigurability of the array geometries, scalability to tens of thousands of qubits, high-fidelity  multi-qubit operations mediated by electronically excited Rydberg states, which can be efficiently controlled using laser fields. 
 
Neutral atom quantum processors also present noise sources that are specific to the platform, including leakage out of the computational subspace through, e.g. spontaneous decay from a highly-excited electronic Rydberg state to a lower-energy state, population remaining in the Rydberg state after a given quantum gate or black-body radiation~\cite{Ma2023}. To address leakage errors, hardware-specific approaches have been suggested~\cite{Scholl2023,Omanakuttan2024} as well as circuit-based techniques~\cite{Cong2022}. For example, in alkaline-earth-like atoms such as $^{171}$Yb, encoding the computational states into  metastable states causes a large fraction of gate errors to leak out of the qubit subspace. These leakage errors can be detected and converted to erasure errors, easier to decode than standard Pauli errors ~\cite{Wu2022,Ma2023,Niroula2024,Yu2024,Omanakuttan2024}. Erasure conversion mechanism is not limited to neutral atom architectures and has also been studied in the context of superconducting circuits~\cite{Suchara2015,Kubica2023,Chang2024,Gu2024,Gu2025} and cold ions~\cite{Kang2023}.

One particularly important source of errors that is specific to the neutral atom platform is the  loss of atoms during computation. Atom loss can occur due to several mechanisms, including heating that causes atoms to escape the optical traps, collisions with background gas particles, or anti-trapping during an excitation to a Rydberg state~\cite{Bluvstein2022,Cong2022,Evered2023,Chow2024,Kobayashi2024}. We note that Rydberg leakage errors can also be converted into atom loss errors using the repulsive force of the trap~\cite{Chow2024}. A promising approach to identify and correct atom losses is provided by so-called loss detection units (LDU)~\cite{Gottesman1997,Preskill1998,Knill2005,Suchara2015,Stricker2020,Cong2022,Moses2023,Chow2024}, which are small circuits attached to each data qubit, entangling the latter with an additional atom. The performances of QEC protocols supplemented with theses LDUs in the presence of atom loss have not been evaluated to date.

In this work, we investigate QEC protocols supplemented with two types of LDUs: a standard LDU~\cite{Gottesman1997,Preskill1998}, which operates by entangling the LDU's atom with the presence of the data qubit, and a teleportation-based LDU scheme~\cite{Knill2005,Chow2024} where the state of the original data atom is transferred to a new, fresh atom. We focus on the rotated surface code~\cite{Bravyi1998,Dennis2002,Fowler2012,Bluvstein2024}, which has emerged as a leading choice for fault-tolerant quantum computation. 
We investigate its tolerance to atom loss errors and depolarizing noise, a common approach for modelling noisy gates, by performing circuit-level simulations.  We derive the scaling laws for logical error probability in the asymptotic regime of low physical error probability.  We demonstrate the existence of an error threshold that depends linearly on the probabilities of both atom loss and depolarizing noise. In particular, for zero depolarizing noise, we find an atom loss threshold around $2.6\%$ for both LDU schemes. Finally, we conduct a comparative analysis of both LDU protocols, finding that the teleportation-based LDU outperforms the standard LDU in terms of achievable logical error probabilities, since it requires fewer two-qubit gates. It also avoids the need for active feedback. Both schemes require approximately the same number of atoms, with a slight advantage for the teleportation-based LDU.

For our analysis, we extend the minimal weight perfect matching (MWPM) decoder~\cite{Edmonds1965,Edmonds1965a,Higgott2021,Higgott2025}, a widely used decoding algorithm, to handle losses. Specifically, we incorporate the available information about the loss locations, namely the rounds and positions of lost qubits, into the algorithm by modifying the graph's weights on which MWPM is executed. In the absence of depolarizing noise, this allows us to correct up to $d-1$ errors for a surface code of distance $d$, analogous to an erasure channel. The decoding algorithm leads to a logical error probability reduction of almost three orders of magnitude compared to a naive decoding scheme.  \\
 
 This article is organized as follows: Section~\ref{sec:noise} outlines the error models used in our study, including both atom loss and depolarizing noise. Section~\ref{sec:protocols} introduces the two distinct LDU protocols — the standard LDU and the teleportation-based LDU — detailing their design and operations. In Section~\ref{sec:decoder}, we present the principle of the new decoder adapted for losses. Using this decoding scheme, we analyze in Section~\ref{sec:logical error} the logical error probabilities for both protocols under varying noise conditions. We further discuss the results and compare the performance of the protocols. To emphasize the performance of our decoding scheme, we compare it to a naive decoder in Section~\ref{sec:perf decoder}. In Section~\ref{sec:sim_method}, we detail the numerical method used to implement the surface code supplemented with these LDUs. Finally, in Section~\ref{sec:real_loss}, we investigate a refined loss model that accounts for the impact on the remaining atom when the other atom is lost during the execution of a CZ gate.
\section{Error models}
\label{sec:noise}
In this section, we detail the circuit-level error models used to evaluate the performance of QEC. In Sec.~\ref{sec:loss}, we first describe the atom loss model which accounts for losses during gate operations due to Rydberg excitation and other mechanisms such as collisions and heating effects. We then present the depolarizing noise model, which characterizes imperfections arising from pulse applications and environmental interactions in Sec.~\ref{sec:depo}. These models are combined to provide an accurate description of errors on neutral atom systems.
\subsection{Atom loss}
\label{sec:loss}

 A common approach to entangle neutral atoms consists in exciting them to their Rydberg states and  engineering a CZ gate through the Rydberg blockade effect \cite{Jaksch2000,Jandura2022}. The latter consists in the excitation of one atom to a high-energy Rydberg state preventing nearby atoms from being similarly excited due to strong van-der-Waals or dipole-dipole interactions~\cite{Morgado2021}. This procedure can result in the anti-trapping of the atom in the optical tweezer while in the Rydberg state, leading to a potential loss. 
 
\begin{figure}
     \centering
     \includegraphics[width=\linewidth]{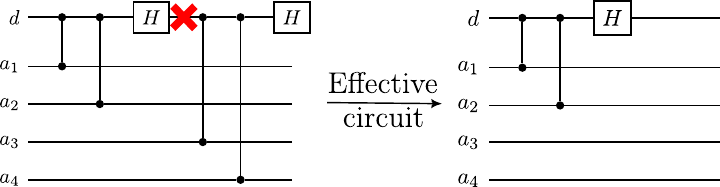}
     \caption{Left circuit shows the typical operations seen by a bulk data qubit during a round. The red cross indicates loss of the data qubit between the first Hadamard gate and the third CZ gate. To simulate the loss, subsequent operations involving the lost atom are removed as shown in the right ("Effective") circuit.}
     \label{fig:effective_circuit}
 \end{figure}

 In our work, we assign to each two-qubit CZ gate a certain probability $p_l$ of losing an atom and assume that if both atoms are lost during the gate, the two losses can be described as two independent events (leading to a joint loss probability of $p_l^2$). In principle, correlated loss—where both atoms are lost with a probability proportional to $p_l$—can occur in real experiments~\cite{Wu2022}. While one might intuitively expect that losing two atoms degrades performance, it is also possible that such correlations could be exploited to enhance decoding. However, we neglect this mechanism here and leave its investigation for future work.
 Furthermore, when an atom is lost during a CZ gate, it is assumed that no entangling operation is carried out on the remaining atom.  Any subsequent gates involving the lost atom (both one-qubit and two-qubit gates) is erased until a new atom has been reloaded (see Fig.~\ref{fig:effective_circuit}). However, as discussed in Sec.~\ref{sec:depo}, two-qubit noise channels are conserved in order to simulate the application of a noisy pulse on the remaining atom.
 
 We note that in principle the remaining atom  may be impacted by the loss of the other atom during the CZ gate, possibly leading to a modified noise channel. In Sec.~\ref{sec:real_loss}, we explore a refined loss model where the remaining atom experiences a maximally biased Z error.
 %In the context of the surface code, we expect that neglecting this latter modification will not result in significant quantitative changes in our simulations. Indeed, if a data qubit is lost, the ancilla will only measure the X or Z parity of three data qubits instead of four. This will make the measurement output completely random ($50\%$ probability for 0 or 1). Adding an additional Pauli error on the ancilla in this scenario does not further impact the measurement probability.

 Other atom loss mechanisms, such as collisions and heating effects, typically occur between gate applications. These mechanisms are incorporated into the loss probability $p_l$ of the subsequent CZ gate, as they produce an equivalent effective circuit (i.e. the same gates are erased).  For simplicity, we assume that the time interval between consecutive CZ gates is roughly uniform, making these additional errors identical for each CZ gate.

 Additionally, we note that, if one of the two atoms is already lost before a given CZ gate is performed, the remaining atom can still be lost since a pulse is applied to it despite the absence of the other atom.
 As a result, the probability of an atom being lost at the $i$-th CZ gate, assuming no fresh atom has been reloaded, is given by
\begin{align}
    p_{l,i}=p_l(1-p_l)^{i-1}
    \label{eq:usual_loss}
\end{align}
where  the term $(1-p_l)^{i-1}$ gives the probability that the atom is not lost in the first $i-1$ CZ gates. A detailed explanation of the simulation method is provided in Sec.~\ref{sec:sim_method}.
\subsection{Depolarizing noise}
\label{sec:depo}
   
 Imperfect pulse applications and coupling to environmental degrees of freedom introduce noise into gates. This noise can be converted into a Pauli noise channel using the randomized compiling method~\cite{Kern2005,Wallman2016,Hashim2021}. To model these imperfections, it is standard to apply a simple depolarizing noise channel (unbiased Pauli noise channel) of error probability $p_d$ after each CZ gate. %Since pulses implementing {\color{red}XXX two-qubit?XXX operations are always applied on the atoms regardless of the presence or absence of the other atoms}, w
 Here, we consider that an atom involved in a CZ gate where the other atom has been lost still experiences a depolarizing noise channel. Combining this standard noise model with the atom loss mechanism allows us to qualitatively assess the performance of the LDU integrated in a QEC protocol as explained below. 

\section{Protocols}
\label{sec:protocols}
 To improve the resilience of the surface code against atom loss, we attach an LDU to each data qubit at the end of each round of error correction (hereafter, the terms "round" and "cycle" are used interchangeably), to verify its presence. 
 
 An LDU is composed of a dedicated ancilla atom coupled to each data atom through a 2-qubit entangling operation and a measurement. In our simulation, we assume that the measurement on the atom is perfect, non-destructive and can distinguish three outcomes: the absence of the atom and the computational states $\ket{0}$ and $\ket{1}$. In experiments, this can typically be realized using a protocol that detects the atom in one of the two computational states (say $\ket{0}$). By performing an X gate and a second measurement, the state $\ket{1}$ can be read out. If no atom has been detected in either of these two measurements, a loss has occurred. We note that alternative approaches involving a low-loss state detection technique~\cite{Fuhrmanek2011,Lis2023,Chow2023,Chow2024} are also possible. 
 
 In this work, we study two versions of LDUs, namely the standard LDU~\cite{Preskill1998,Cong2022,Chow2024} and the teleportation-based LDU~\cite{Knill2005,Chow2024}, as presented in Secs.~\ref{sec:standLDU} and ~\ref{sec:telLDU} below, respectively. We further note that the need for an LDU is restricted to data qubits only, as they cannot be directly measured  without compromising the logical quantum information. Ancilla qubits used for stabilizer measurements do not require LDUs, as their presence (or absence) can be directly inferred from their measurements performed at the end of each cycle of error correction.
 
\begin{figure*}[t]
    \centering
    \includegraphics[width=0.9\linewidth]{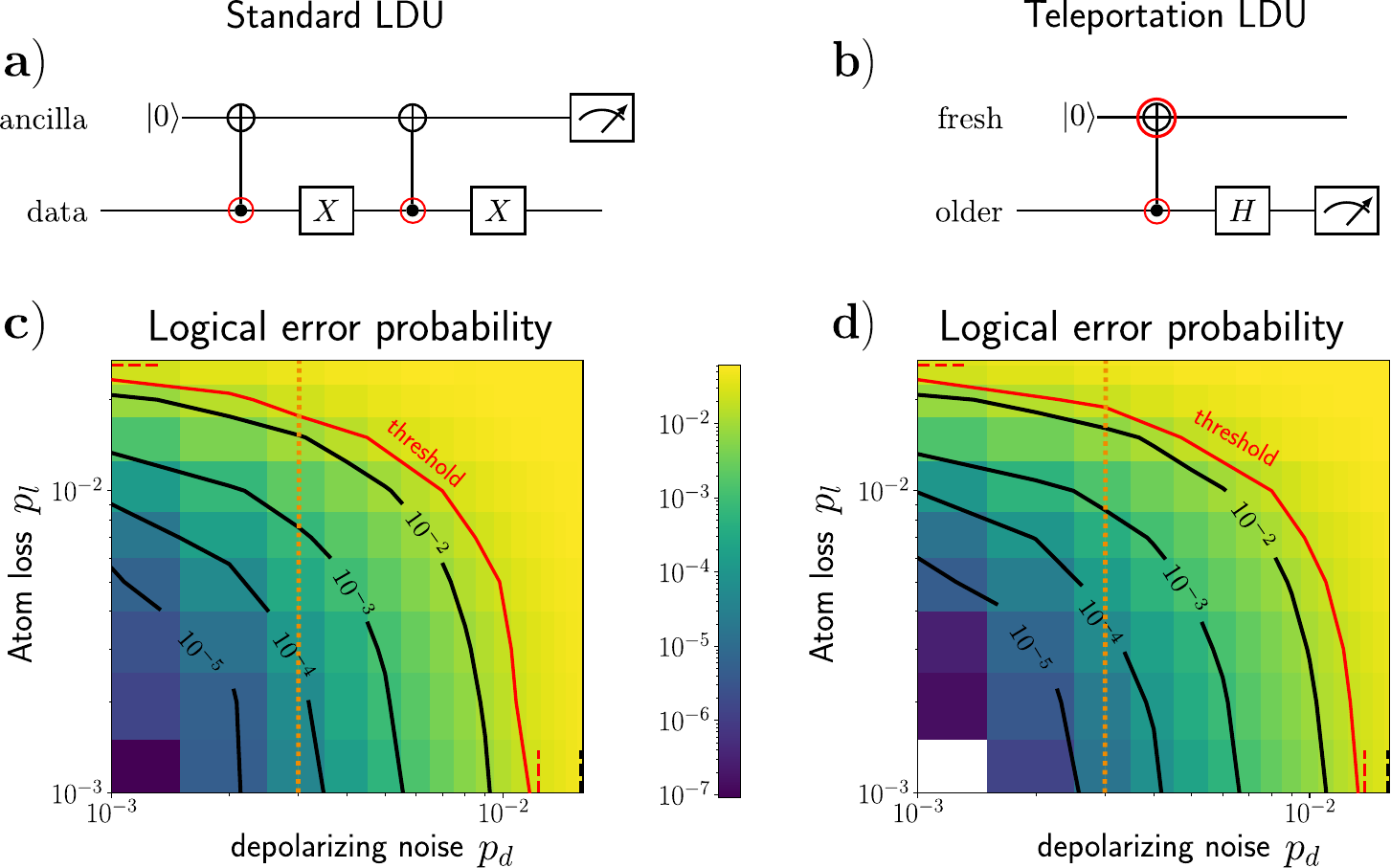}
        \caption{Top: Circuit representations of the two LDU protocols where ${\bf a)}$ shows the standard LDU and ${\bf b)}$ the teleportation LDU. In each cycle, data qubits are coupled to ancilla qubits using one of these protocols. Red circles indicate the potential locations where atoms might be lost in the LDU. The loss of the ancilla qubit in the standard LDU can always be detected using the measurement scheme, allowing for another LDU to be reapplied if necessary. This effectively prevents any loss on them.
        Bottom: The logical error probability normalized by the number of rounds for a surface code of distance $d=11$ and $d$ rounds of stabilizer measurements as a function of the loss probability $p_l$ and the depolarizing error probability $p_d$. The initial state is prepared in $\ket{0}_L$. ${\bf c)}$ corresponds to the standard LDU protocol, and ${\bf d)}$ represents the teleportation-based LDU protocol. Each logical error probability was evaluated with $10^5$ shots, except for regions with $p_l<4. 10^{-3}$ and $p_d<2. 10^{-3}$, where $10^6$ shots were employed. The solid red line marks the error threshold while the solid black lines show curves of constant logical error probability.  The horizontal and vertical dashed red lines indicate the threshold at vanishing depolarizing noise and loss probability, respectively. For comparison, the threshold of the standard surface code is shown by a black dashed line. In {$\bf d)$}, the blank region indicates no errors were found.
        The dashed orange line indicates the current best CZ gate fidelity achieved in neutral atom systems, corresponding to a depolarizing error probability of $p_d=0.003$. A plot of the slice 
 is shown in Fig~\ref{fig:depo_0.003}. 
        }
        \label{fig:error_loss_depo}
\end{figure*}
\subsection{Standard LDU}
\label{sec:standLDU}

 In the standard LDU protocol (see Fig.~\ref{fig:error_loss_depo}${\bf a}$), the presence or absence of data qubit is entangled with the computational states of the ancilla atom. This is achieved by performing two CNOT gates where the data atom is the control qubit and the ancilla is the target qubit, with an X gate applied to the data qubit in between the CNOT gates to ensure that a bit-flip operation is carried out on the ancilla regardless of the control qubit's state. If the data atom is absent, no operation is performed on the ancilla, which remains in its initial state. In that case, a new atom is reloaded in place of the data qubit and initialized in $\ket{0}$.

 LDU's CNOT gates can be recompiled into CZ gates that are natively realized in neutral atom quantum processors by introducing Hadamard rotations on the target qubits. As a result, atom losses associated with the CZ gates can also occur during the execution of the LDU. If the ancilla is lost, its absence is detected by its measurement, and a new ancilla atom is introduced to perform another LDU. However, if the data qubit is lost during the second CZ gate of the LDU, the loss has $50\%$ probability to not be detected, resulting in an atom not being reloaded in the next cycle.

 Additionally, introducing depolarizing noise to the CZ gates renders the LDU imperfect:
 $i)$ First, an effective 1-qubit depolarizing noise channel of error probability $p_{d_1}^{\text{tel. LDU}}$ (see Sec.~\ref{sec:sim_depo}) is induced on the data atom.
 $ii)$ Second, with probability $p_\text{flip}$ (see Sec.~\ref{sec:sim_depo}), the LDU might falsely indicate the absence of atoms while still present or fail to detect the loss of an atom causing the atom to remain absent potentially over multiple rounds.

\subsection{Teleportation-based LDU}
\label{sec:telLDU}
 In the teleportation-based LDU~\cite{Knill2005,Chow2024} (see Fig~\ref{fig:error_loss_depo}${\bf b}$), data qubits are systematically replaced by fresh atoms. The state of the older atom is transferred to the fresh one by means of a teleportation protocol using only one entangling gate. During this protocol, the older data qubit is measured, testifying of its presence. There are three possible outcomes:
 
$i)$ When the data atom is present, there is a $50\%$ probability that the state measured is $\ket{0}$, indicating that the state has been perfectly teleported to the new atom.
 
$ii)$ If the measurement output is one, the state of the fresh atom will have a Z (phase flip) error. This Z error is not corrected immediately using a measurement feedback gate; instead, it can be exactly addressed during the decoding process by flipping the surrounding X stabilizer measurements.

$iii)$ In the case where the data atom has been lost, this LDU simply adds a new atom into the circuit initialized to $\ket{0}$. If the fresh atom is lost during the LDU, it cannot be immediately detected, and the next round will proceed without it. It will eventually be detected and a new atom reloaded into the system in the subsequent cycle, even in the presence of depolarizing noise.
 
 Simulating the teleportation LDU with depolarizing noise on the CZ gate induces an effective 1-qubit depolarizing noise channel on the teleported state of error probability $p_{d_1}^{\text{tel. LDU}}$ (see Sec.~\ref{sec:sim_depo}), resulting in incorrect indications for the heralded Z error. \\

\subsection{Qualitative comparison of both protocols and atom consumption} 

 The teleportation-based LDU offers several advantages over the standard protocol. First, it requires only one CZ gate compared to two in the standard protocol. Second, it eliminates the need for active feedback gate during computation, whereas the standard protocol necessitates different actions depending on the ancilla's state. Finally,  replacing all data qubits every round helps to cool down the system, thereby reducing losses due to heating.  Although this cooling advantage is not considered in our simulation, it remains a significant benefit~\cite{Chow2024}.

Assuming the ancilla LDU for the standard LDU protocol and the older data atom for the teleportation LDU can be reused in subsequent rounds (provided they are not lost), the atom consumption of the teleportation LDU exhibits a slight advantage over the standard LDU.
Both protocols require trapping $d^2$ additional atoms (one for each data qubit) compared to the standard surface code  (i.e. without LDU).  During each error correction cycle, an average of  $p_{l,\text{cycle}} \times n_\text{atoms}$ atoms are lost, where 
$p_{l,\text{cycle}} =1-(1-p_l)^{n_\text{CZ}}\simeq p_l \times n_\text{CZ}$ represents the probability of losing an atom within a single round. The number of CZ gates experienced by an atom depends on its position in the surface code. For simplicity and to provide an estimate of atom consumption, we focus on atoms in the bulk, which are far more numerous, and neglect those located on the edges.

In both protocols, bulk ancilla atoms used for stabilizer measurements experience four CZ gates per cycle, while bulk data atoms experience six CZ gates per cycle. For the teleportation LDU, fresh atoms are not replaced in the same round as the older ones if they are lost. Thus, the atom loss only accounts for older data atoms, where we take into accounts the LDU operations acting on the data qubit during both the previous and current rounds. Consequently, the average atom consumption per cycle for the teleportation LDU scales as:
 \begin{align}
     n_{\text{tel. LDU}}\sim 10 \times p_l\times d^2 
 \end{align}
 Additionally, the standard LDU requires considering the replacement of the ancilla LDU. They experience approximately two CZ gates per round (neglecting that we reapply a new LDU whenever the ancilla LDU is lost). As a result, the average number of atoms that need to be reloaded per cycle for the standard LDU scales as:
 \begin{align}
     n_{\text{stand. LDU}}\sim   12 \times p_l \times d^2 .
 \end{align}
 
\section{``Loss-aware'' decoder}
\label{sec:decoder}
In this section, we present a scheme for a ``loss-aware'' decoder which leverages both stabilizer measurement results and LDU measurement outputs to determine the most probable error patterns.  A comparison with a naive decoding scheme that does not incorporate information on the loss locations is conducted in Sec.~\ref{sec:perf decoder}.\\
We note that several approximations are introduced during the decoding procedure to reduce the decoder latency. The latter are used exclusively by the decoder and are never employed in the simulations when sampling Pauli and loss errors (see Sec.~\ref{sec:sim_method} for details).

Our decoder is based on the MWPM algorithm~\cite{Edmonds1965,Edmonds1965a,Higgott2021,Higgott2025} to decode errors resulting from depolarizing noise and atom loss. Error mechanisms generate a detector error model (DEM) consisting of a set of lists of flipped stabilizers along with their probabilities of occurrence. Using Pymatching~\cite{Higgott2021,Higgott2025}, a Python library that implements the MWPM algorithm, a graph is generated where nodes are the stabilizers and the edges represent error mechanisms pairing the corresponding flipped stabilizers. Each edge is weighted by the log-likelihood ratio $\text{ln}(\frac{1-p}{p})$ where $p$ is the probability of the error~\cite{Dennis2002,Higgott2021,Higgott2025}. This graph is extended in the time direction to account for multiple rounds of error correction, enabling us to decode errors stemming from stabilizer measurements~\cite{Fowler2012}.
\begin{figure}[!!h]
    \centering
    \includegraphics[width=0.9\linewidth]{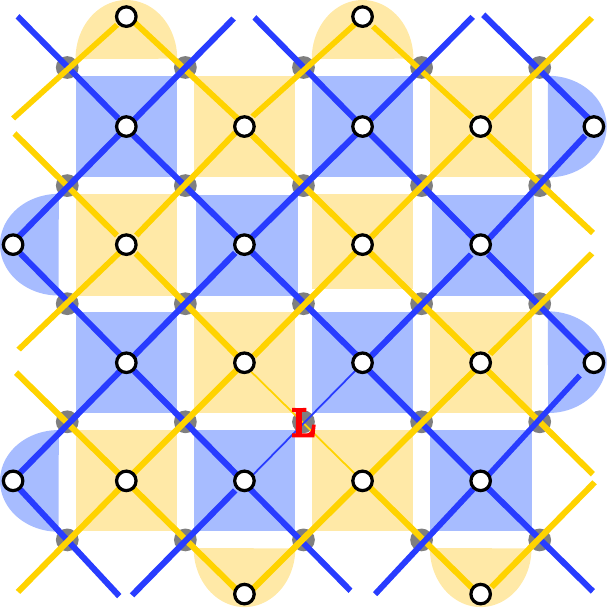}
    \caption{A schematic representation of the "loss-aware" DEM. Orange (respectively blue) semi-transparent plaquettes correspond to $Z$ (resp. $X$) stabilizers, which detect $X$ (resp. $Z$) errors. Black semi-transparent dots represent data qubits. Overlaid on the surface code is the DEM, consisting of orange (resp. blue) edges connecting neighboring ancilla qubits, representing $X$ (resp. $Z$) errors occurring on the intermediate data qubit, typically due to depolarizing noise. The edge thickness reflects their weights. In the case of data qubit loss (denoted by a red $L$) detected by its LDU, the weights of edges connecting neighboring stabilizers are reduced, as detailed in the main text. The figure illustrates a single time slice corresponding to one cycle of quantum memory operation. To fully capture the error dynamics, the model must be extended to the time domain to include errors on ancilla qubits.   }
    \label{fig:loss dem}
\end{figure}
At each round, stabilizer measurements produce a syndrome, which is a list of flipped stabilizers. The MWPM decoding procedure consists in pairing these flipped stabilizers by finding the set of paths that minimizes the sum of the edge weights. This corresponds to identifying the most likely error pattern that explains the observed syndrome assuming uncorrelated X and Z errors — thus approximating Y errors as simultaneous, independent X and Z errors.

Stim~\cite{Gidney2021}, the Clifford simulator used to evaluate the performance of the surface code, automatically generates the DEM stemming from the depolarizing noise channel. We adapt the decoder to include errors arising from atom loss by updating the graph's weights based on the cycles where losses have been detected. This allows us to take advantage of the additional information given by the LDU, making the decoding more precise (for a sketch, see Fig.~\ref{fig:loss dem}). The idea is similar to~\cite{Sahay2023} where an adapted version of MPWM was used to address erasure errors; however, in that case, the exact location of the error was known, whereas, in our scheme, only the cycle is identified.

To achieve this, we augment the DEM generated by the depolarizing noise channel with new error mechanisms arising from atom loss. The absence of an atom at the end of a round $r$ results in a list of potential loss locations. Each potential loss location generates its own DEM. The probabilities of these DEMs are then weighted by the likelihood that the loss occurred at that specific location, conditioned that a loss was detected in the round $r$. These weighted DEMs are subsequently added to the final DEM, allowing the decoder to account for both depolarizing noise and atom loss in the error correction process. Our decoder is similar to the one developed in~\cite{Suchara2015} for the case of a SWAP-based LDU and a different loss model. Our approach offers the advantage of being easily automated using the DEM method implemented in Stim.

We determine the conditional probability for each loss location by computing the probability that the loss occurred at that specific location $p_{l,i}$, normalized by the total probability of all possible loss locations $p_{\text{tot.},r}$

\begin{align}
p_{\text{pot.loss},r,i}=\frac{p_{l,i}}{p_{\text{tot.},r}}
\label{eq:cond_proba}
\end{align}
where $r$ is the round where the loss has been detected and $i$ is the position of the CZ gate within the round. In Sec.~\ref{sec:loss_proba}, loss probabilities $p_{l,i}$ are explicitly computed. In Appendix ~\ref{ap:decoding}, we calculate these conditional probabilities for the different cases (loss of a stabilizer ancilla atom or of a data atom and type of LDU used).\\

In this procedure, we make the approximation that the propagation of depolarizing errors in the circuit is unaffected by atom loss, except in the case of stabilizer ancilla loss, as the latter leads to the absence of a node in the weighted graph. To handle this, we reconnect all edges that were linked to the lost node to the node representing the same stabilizer but at the subsequent cycle of error correction. If there is no subsequent cycle, the edges are paired to the vacuum.

Furthermore, when using Stim to construct the circuit, one must determine whether a stabilizer has flipped by comparing its measurement outcome at a given round to the result from the previous round — this is referred to as a {\it detector}. If a stabilizer ancilla atom is lost, the measurement result in the next round of error correction cannot be compared to the previous one. To resolve this, we instead compare it to the most recent round prior to that loss.

Finally, the edge weights of the final DEM are defined as the sum of the independent loss events. This approach makes our decoder quite efficient, as we can precompute and store the DEM for each potential loss location before running the simulation. During the decoding process, we simply retrieve these precomputed DEMs whenever a loss at the corresponding location is detected.

Although the decoding procedure is based on the MWPM algorithm, it can be readily adapted for other decoders utilizing the DEM, such as Belief Propagation~\cite{Kschischang2001}. This is because the procedure primarily involves generating a suitable DEM, with only the final step requiring the conversion of the DEM into a graph-like structure.

\section{Logical error probability}
\label{sec:logical error}
 In this section, we analyze the logical error probability of the rotated surface code in the presence of depolarizing noise and atom loss for different code distances $d=3,5,7,9,11$ and $d$ cycles of error correction. This analysis allows us to determine the threshold for various combinations of depolarizing error probabilities $p_d$ and loss probability $p_l$. We benchmark the performance of both LDU protocols (see Fig.~\ref{fig:error_loss_depo}) discussed in Sec.~\ref{sec:protocols}: the standard LDU and the teleportation-based LDU.

 In Fig.~\ref{fig:error_loss_depo}${\bf c}$,${\bf d}$, we present the logical error probability of an initial state prepared in $\ket{0}_L$ for a code distance of $d=11$, normalized by the number of rounds $d$ as $\varepsilon_r$: $\varepsilon_r=1-(1-\varepsilon)^{1/d}$, where $\varepsilon$ is the total logical error probability. This normalization allows for a fair comparison across various code sizes by computing a quantity independent of the numbers of rounds~\cite{Acharya2023,Xu2024,Pecorari2025}. The plot shows the logical error probability at various depolarizing error probabilities $p_d$ and loss probabilities $p_l$. The threshold is indicated by a solid red line. The  red dashed vertical (resp. horizontal) line represents the threshold at vanishing loss probability (resp. depolarizing noise). For comparison, the black dashed vertical line shows the threshold for the standard surface code (i.e. without LDU and atom loss). The logical error probability for an initial state in the $\ket{+}_L$ is presented in Appendix~\ref{ap:xbasis} and gives similar behavior.

\subsection{Vanishing depolarizing noise}
\begin{figure}[t] 
%\begin{figure}[!!h]
%\begin{figure}[t]
\begin{tabular}{l}\includegraphics[width=0.9\linewidth]{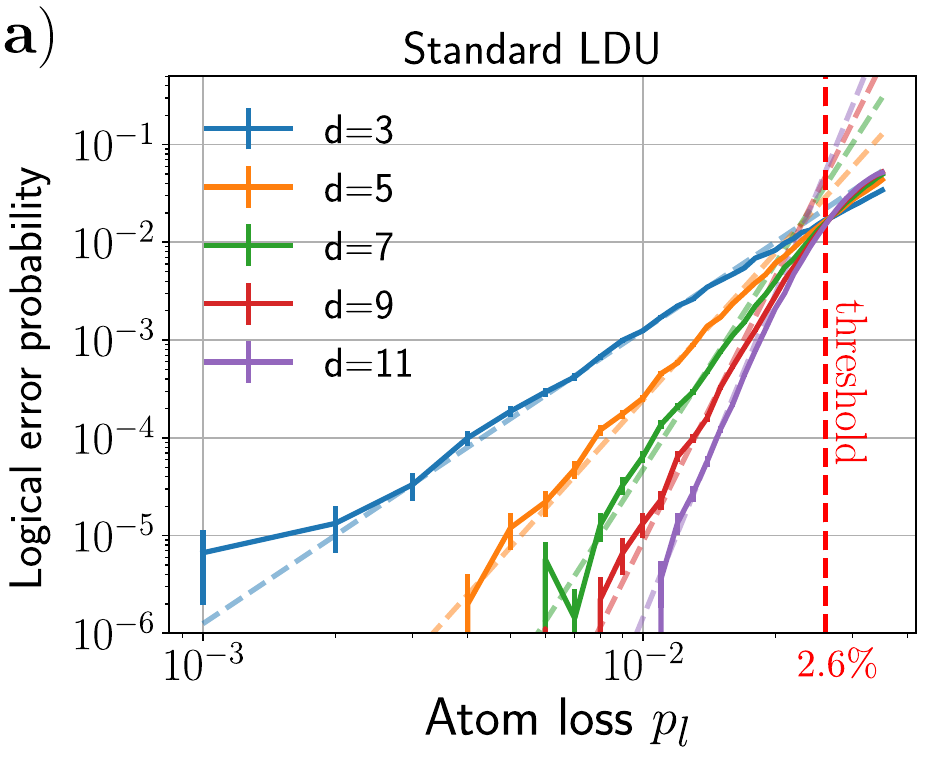} \\  \includegraphics[width=0.9\linewidth]{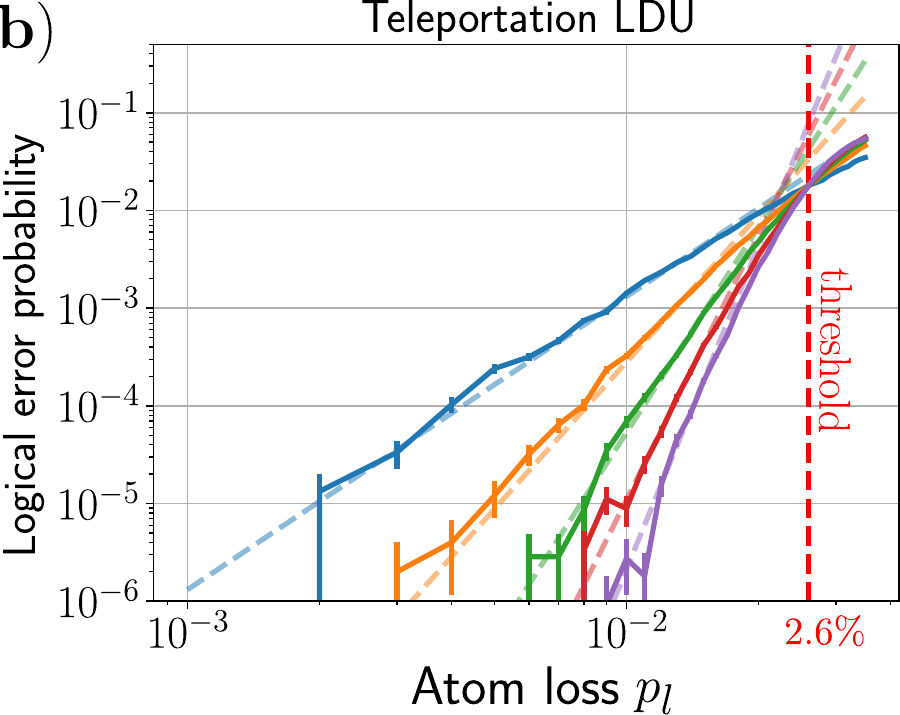}
\end{tabular}
    \caption{Logical error probability normalized by the number of rounds 
 at vanishing depolarizing noise as a function of the loss probability for code distance $d=3,5,7,9,11$ and $d$ cycles of stabilizer measurements for ${\bf a)}$ the standard LDU protocol and ${\bf b)}$ the teleportation LDU protocol. The dashed red vertical line indicates the threshold at $2.6\%$ for both protocols. Other colored dashed lines represent fits of power laws $d$ for each curve respectively. $10^5$ shots were used to estimate the logical error probability.}
    \label{fig:no depo}
\end{figure}
First, we examine the performance of the LDU protocol in the absence of depolarizing noise. The LDU heralds the cycle in which the atom is lost, allowing this error mechanism to behave like an erasure channel. A key result is the scaling of the logical error probability as a power law as a function of the physical error probability with an exponent $d$~\cite{Nielsen2010}. This expectation is confirmed by the data shown in Fig.~\ref{fig:no depo}, where we plot the logical error probability as a function of the loss probability $p_l$ (solid lines). The figure shows a good fit between numerical data and the power-law function of exponent $d$  (dashed lines) for all cases, as expected.

In the absence of depolarizing noise, we observe that both protocols yield the same logical error probability (within error bars) as well as the same threshold approximately at $2.6\%$ (see Fig.~\ref{fig:no depo}). This is due to the fact that, although the standard LDU has two CZ gates and the teleportation-based LDU only one, both schemes exhibit the same number of potential loss locations (see red circles in Fig.~\ref{fig:error_loss_depo}{\bf a,b}): In the standard LDU, the loss of an ancilla can be detected from its measurement, allowing for the reapplication of another LDU. Consequently, only the loss of data qubits matters. On the contrary, in the teleportation-based LDU, the loss of the fresh atom cannot be detected within the same cycle of error correction. Therefore, two loss locations are possible in both LDUs:  the data qubit can be lost in either of the two CZ gates in the standard LDU, or one of the two atoms involved in the teleportation-based LDU can be lost during the single CZ gate.

\subsection{Vanishing atom loss probability}

  \begin{figure}[!!h]
  %\begin{figure}[t]
\includegraphics[width=0.9\linewidth]{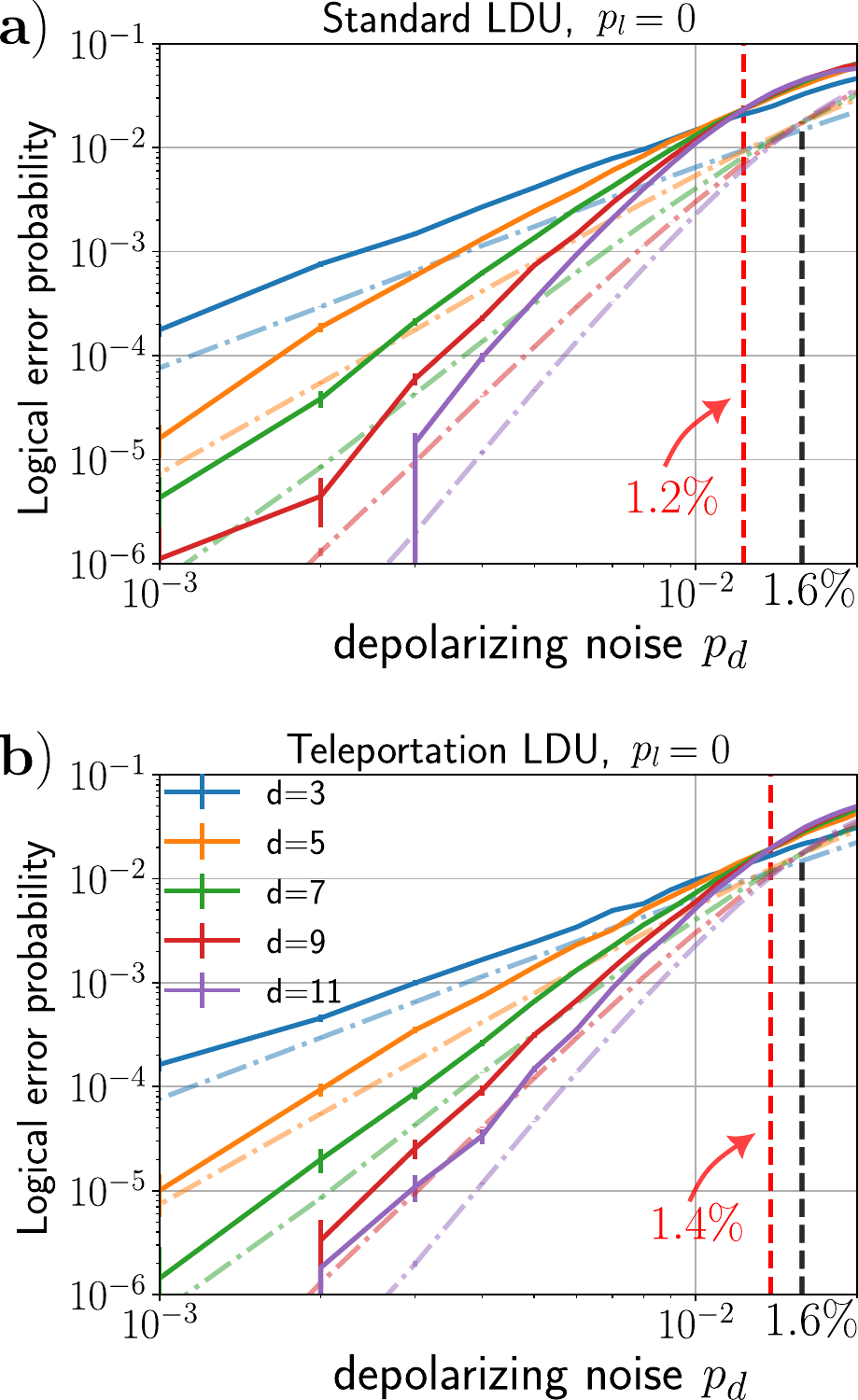}  
    \caption{The logical error probability, normalized by the number of rounds at vanishing loss probability, is shown as a function of depolarizing noise error probability $p_d$ for code distances $d=3, 5, 7, 9, 11$ and $d$ cycles of stabilizer measurements. The plot includes results for ${\bf a)}$ the standard LDU protocol and ${\bf b)}$ the teleportation LDU protocol. The red dashed vertical line marks the threshold at $1.2\%$ for the standard LDU protocol and $1.4\%$ for the teleportation LDU protocol. For comparison, semi-transparent dashed-dotted lines represent the logical error probability per round for the standard surface code (i.e. without LDU), with its threshold indicated by a black dashed line. The logical error probability is estimated using $10^5$ shots. }
    \label{fig:no loss}
\end{figure}

 In the limit of a vanishing loss probability ($p_l=0$) and finite depolarizing noise error probability $p_d$, instead, one recovers the scaling for the logical error probability vs $p_d$ typical of the standard surface code without LDU, which is characterized by a power law with exponent $\frac{d+1}{2}$ (see Fig.~\ref{fig:no loss}). The logical error probabilities of the LDU protocols are higher and their thresholds are lower than that of the standard surface code (approximately $1.6\%$) due to the additional CZ gates used in the LDU. Under these conditions, the teleportation-based protocol achieves a lower logical error probability and a slightly higher threshold, around $1.4\%$, compared to the standard LDU protocol, which has a threshold of approximately $1.2\%$, as the teleportation protocol uses only one CZ gate per LDU.

 \subsection{Finite depolarizing noise errors and atom loss probabilities}

In Fig.~\ref{fig:error_loss_depo}, we plot the logical error probability with respect to the loss probability $p_l$ and the depolarizing noise error probability $p_d$. A major result of this computation is the existence of an error threshold (see red line) below which scaling up the system exponentially suppress the logical errors. The logical error probability is primarily influenced by the scaling of the depolarizing noise, $\frac{d+1}{2}$. By plotting the logical error probability as a function of both the loss probability $p_l$ and the depolarizing noise error probability $p_d$ with linear scale axes in Fig.~\ref{fig:linear} in Appendix~\ref{ap:linear}), we find that iso-logical error probability curves i.e. curves with constant logical error probability are approximately straight lines. Motivated by this observation, we conjecture that the logical error probability  can be generally described in the asymptotic low-error regime by a function of the form
 \begin{align}
    f(p_l,p_d)=(\alpha_d p_d+\beta_d p_l)^{\frac{d+1}{2}}
    \label{eq:fit1}
\end{align}
where $\alpha_d$, $\beta_d$ are two fitting parameters, which depend on the distance code $d$.
To ensure consistency with the power law scaling of $d$ at vanishing depolarizing noise, we introduce counterterms to Eq.~\eqref{eq:fit1}, resulting in the following proposed ansatz function for the dependence of the logical error probability on $p_l$ and $p_d$ 
\begin{align}
    f(p_l,p_d)=(\alpha_d p_d+\beta_d p_l)^{\frac{d+1}{2}}-(\beta_d p_l)^{\frac{d+1}{2}}+(\gamma_d p_l)^{d}
    \label{eq:fit2}
\end{align} with $\gamma_d$ a third fitting parameter.\\

\begin{figure}[t]
    \centering
    \includegraphics[width=0.9\linewidth]{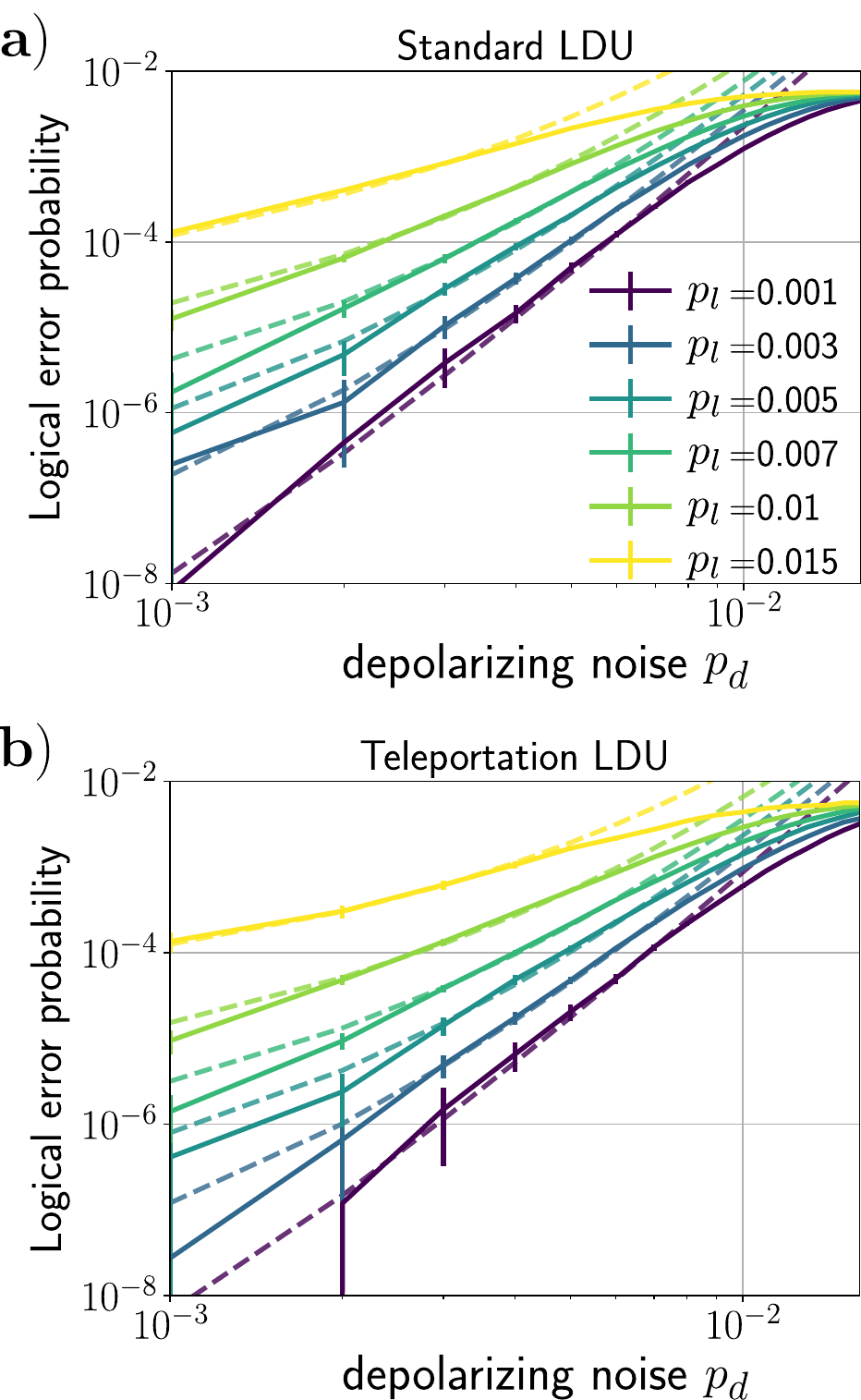}
    \caption{Logical error probability normalized by the number of rounds 
 for various loss probabilities $p_l$ as a function of the depolarizing error probability $p_d$ for a code distance $d=11$ and $d$ cycles of stabilizer measurements. ${\bf a)}$ represents the standard LDU protocol, and ${\bf b)}$ the teleportation LDU protocol. The colored dashed lines correspond to the same fit for the various curves using Eq.~\eqref{eq:fit2}.}
    \label{fig:fit}
\end{figure}

In Fig.~\ref{fig:fit}, we plot example results for the logical error probability normalized by the number of rounds for a distance code $d=11$ as a function of the depolarizing error probability $p_d$ and for several different loss probabilities $p_l$ (solid lines). We fit the latter quantity using the ansatz function from Eq.~\eqref{eq:fit2} (dashed lines). The fitting was performed in the low-error regime, specifically for logical error probabilities per round below $3\times 10^{-4}$. Using three fitting parameters, we successfully reproduce the logical error probability (within error bars) for sufficiently low depolarizing noise and loss probabilities. Discrepancies observed at $p_d=10^{-3}$ are likely due to shot noise simulation. The fitting parameters were found to be approximately $(\alpha_d^\text{stand.},\beta_d^\text{stand.},\gamma_d^\text{stand.})\simeq(35,14,24)$ for the standard LDU and $(\alpha_d^\text{tel.},\beta_d^\text{tel.},\gamma_d^\text{tel.})\simeq(30,14,26)$ for the teleportation LDU.

Despite the counterterms added, we note that Eq.~\eqref{eq:fit2} still describes approximate straight iso-logical error probability lines as $(\alpha_d p_d+\beta_d p_l)^{\frac{d+1}{2}}\gg (\beta p_l)^{\frac{d+1}{2}},(\gamma p_l)^{d}$ for $d=11$ and $0.001\leq p_l,p_d \lesssim 0.02$. 

The first term on the right-hand side of Eq.~\eqref{eq:fit2} indicates that logical errors arise when the total number of errors
 is at least $\frac{d+1}{2}$. Hence, a logical error can occur even if each error source, when considered independently, produces fewer than $\frac{d+1}{2}$ physical errors. An example of such a mechanism with $\frac{d-1}{2}$ depolarizing errors and one atom loss is shown in Appendix~\ref{ap:combined errors}. 
 
Notably, we observe that the threshold lies in the iso-logical error probability curve around $2\%$, therefore,  it depends linearly on the depolarizing noise error probability $p_d$ and atom loss $p_l$.\\

\begin{figure}[!!h]
    \centering
    \includegraphics[width=0.9\linewidth]{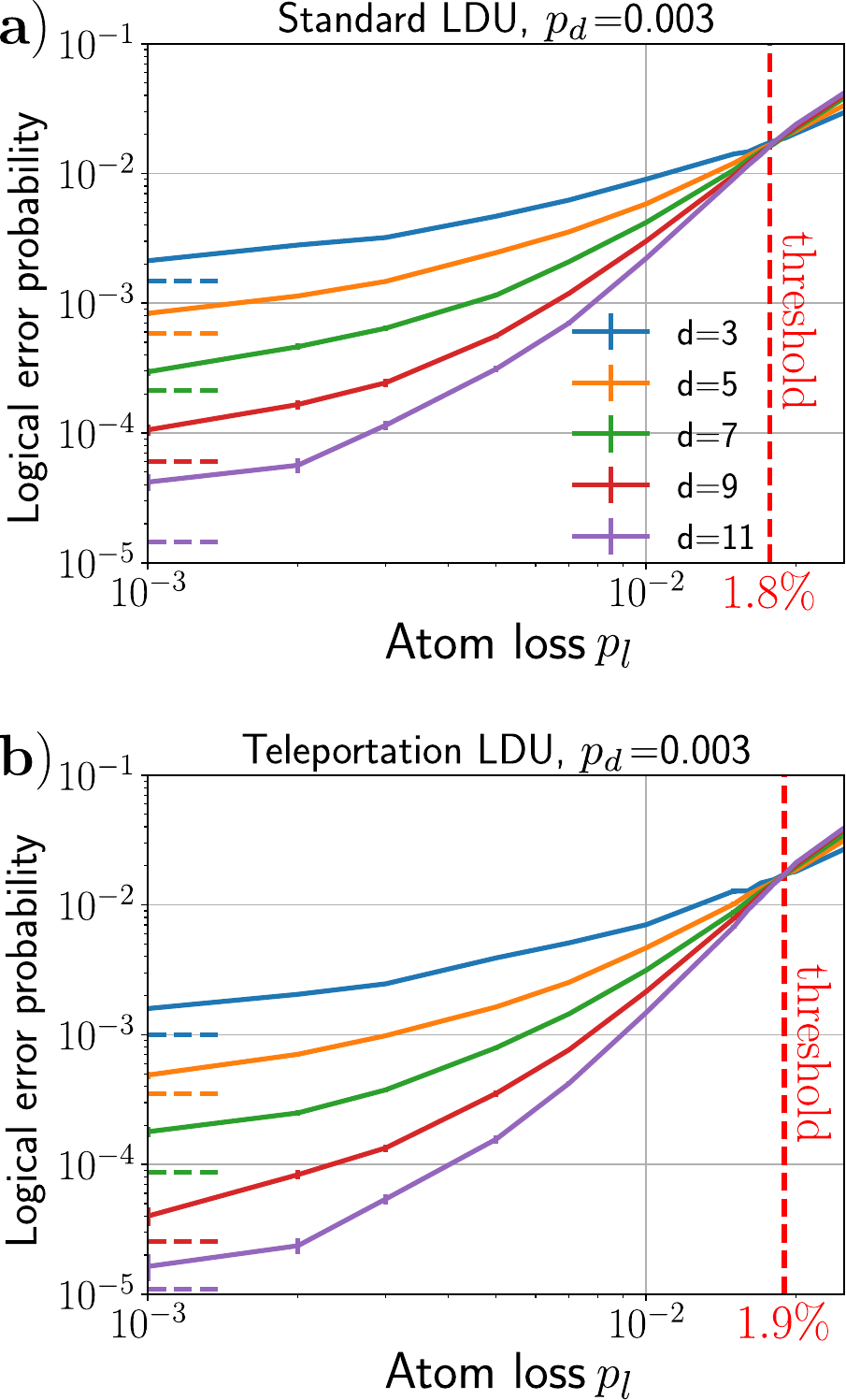}
    \caption{Logical error probability normalized by the number of rounds 
 at fixed depolarizing noise error probability $p_d=0.003$ as a function of the loss probability for code distances $d=3,5,7,9,11$ and $d$ cycles of stabilizer measurements for ${\bf a)}$ the standard LDU protocol and ${\bf b)}$ the teleportation LDU protocol. The dashed red vertical line indicates the threshold at $1.8\%$ (resp. $1.9\%)$ for the standard (resp. teleportation) LDU protocol. Horizontal colored dashed lines indicate the value at vanishing loss probability for each curve, respectively. $10^5$ shots were used to estimate the logical error probability.}
    \label{fig:depo_0.003}
\end{figure}

Recently, state-of-the-art experiments have achieved a  Pauli noise error around $0.3\%$ for CZ gates ~\cite{Tsai2025,Radnaev2024, Reichardt2024}. In this context, we plot in Fig.~\ref{fig:depo_0.003} the logical error probability as a function of the loss probability, with the depolarizing noise error probability fixed at $p_d=0.003$. In this regime of parameters, our results indicate that the teleportation LDU protocol performs slightly better than the standard LDU protocol: The atom loss threshold occurs at $1.8\%$ for the standard LDU and at a comparable $1.9\%$ for the teleportation LDU, which however also achieves a lower logical error probability. 

We note that atom losses and leakages may vary depending on the atomic platform and has been estimated from $0.02\%$ in~\cite{Tsai2025} up to $1\%$ in~\cite{Bluvstein2022}. While here we consider only the effects of depolarizing noise and atom loss, our numerical simulations in Fig.~\ref{fig:depo_0.003} suggest that current experiments may already be exploring the regime of parameters below threshold, where quantum error correction is possible, in qualitative agreement with recent experimental results~\cite{Bluvstein2024}.\\

\subsection{Comparison of both LDU protocols}
\begin{figure}
    \centering
    \includegraphics[width=\linewidth]{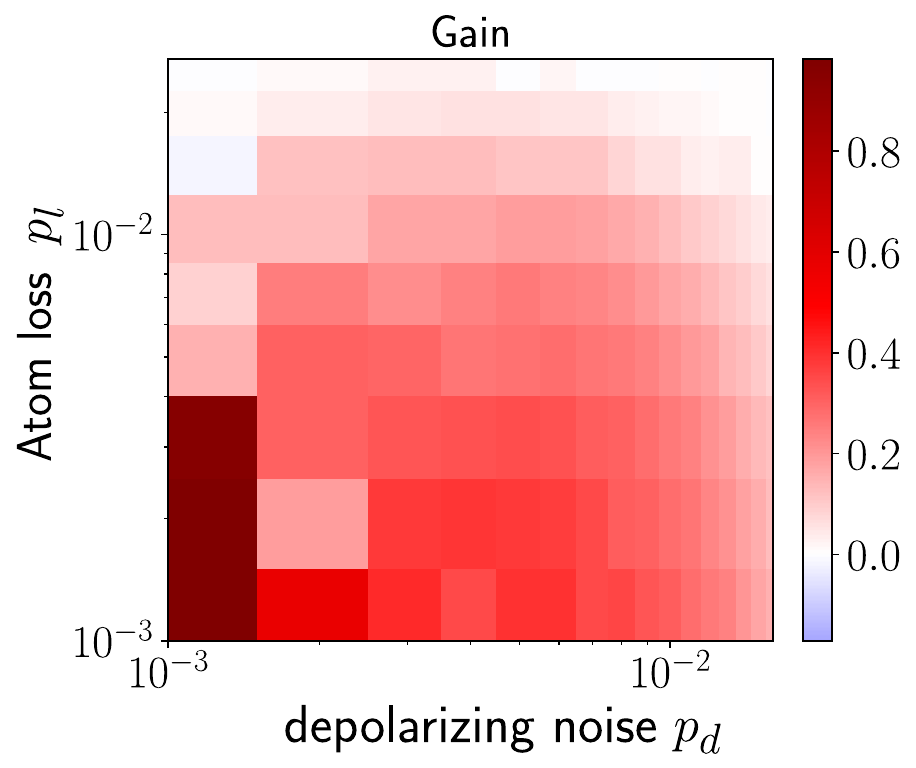}
    \caption{Plot showing the gain $G$ of the teleportation-based LDU protocol compared to the standard LDU protocol as a function of the loss probability $p_l$ and the depolarizing noise error probability $p_d$ for a surface code of distance $d=11$.}
    \label{fig:gain}
\end{figure}

 To further quantify the relative performance of the teleportation-based LDU protocol vs the standard LDU,  in Fig.~\ref{fig:gain} we plot the gain $G$ as a function of the depolarizing noise error probability $p_d$ and the loss probability $p_l$, where $G$ is defined as
 \begin{align}
    G =\text{log}_{10}\left(\frac{\varepsilon_{r,\text{stand. LDU}}}{\varepsilon_{r,\text{tel. LDU}}} \right).
    \label{eq:gain}
 \end{align}
 Here, $\varepsilon_{r,\text{stand. LDU}}$, $\varepsilon_{r,\text{tel. LDU}}$ stand for the logical error probability normalized by the number of rounds of the standard LDU protocol and the teleportation LDU protocol, respectively. The plot clearly shows that, below the error threshold, the teleportation protocol should be favored if atom resources are not a constraint. In fact, for a code of distance $d=11$, switching to the teleportation LDU protocol can already reduce the logical error probability by up to an order of magnitude. Using Eq.~\eqref{eq:fit2} along with the fitting parameters determined in the previous section, we expect the advantage of the teleportation LDU  to be even more pronounced for larger distance code $d$ as $\alpha_d^{\text{stand.}}>\alpha_d^{\text{tel.}}$ and $\beta_d^{\text{stand.}}=\beta_d^{\text{tel.}}$. This represents a key result of our study.

 We note that the value of $G$ presented in Fig.~\ref{fig:gain} for the smallest physical error probabilities $(p_d,p_l)\simeq (10^{-3},10^{-3})$ is affected by poor statistics: In fact, in this regime $\varepsilon_{r,\text{tel. LDU}}=0$ due to the limited number of shots performed in the numerics, which would result in an infinite gain. To avoid this, we substitute the measured $G$ for $(p_d,p_l)\simeq (10^{-3},10^{-3})$ with a lower bound $G\simeq1$, corresponding to the maximum $G$ measured across the entire plot.

\section{``Loss-aware'' vs. naive decoding performance }
\label{sec:perf decoder}
To quantify the effectiveness of the ``loss-aware'' decoder introduced in this work, we compare its performance with that of a naive decoding scheme. In the naive approach, decoding is also based on the generation of a graph on which the MWPM algorithm is applied; however, its weights depend solely on the depolarizing noise channel error probability and the prior knowledge of the loss probability, without incorporating information on the loss locations.  This approach does not update weights in response to detected losses.

%Consequently, losses are not treated as an erasure channel. Thus, in the regime without any depolarizing noise but only losses, the logical error probability follows a power law of exponent $\frac{d+1}{2}$ (as opposed to $d$ for the ``loss-aware'' decoder, see Fig.~\ref{fig:no depo}). This is shown in Fig.~\ref{fig:naive no depo} where the logical error probability normalized by the number of rounds for both LDU protocols is plotted accross various code distances (see solid lines). Fits with the $\frac{d+1}{2}$ power law are performed (see dashed lines) for the different code sizes. Notably, the loss threshold at $ $ is lower than that of the ``loss-aware'' decoder ($2.6\%$), and the logical error probability is higher for a given code distance. 

Similarly to Eq.~\eqref{eq:gain},  we define the gain of the ``loss-aware'' decoder over the naive one for a given LDU protocol as
\begin{align}
    G_\text{LDU}^\text{decoder}=\text{log}_{10}\left(\frac{\varepsilon_{r,\text{LDU}}^{\text{naive}}}{\varepsilon_{r,\text{LDU}}^{\text{loss-aware}}} \right),
    \label{eq:gain decoder}
\end{align}
where $\varepsilon_{r,\text{LDU}}^{\text{naive}}$ (resp. $\varepsilon_{r,\text{LDU}}^{\text{loss-aware}}$) is the logical error probability normalized by the number of rounds for a given LDU protocol obtained using the naive (resp. ``loss-aware'') decoder.

\begin{figure}[!!h]
    \centering
    \includegraphics[width=\linewidth]{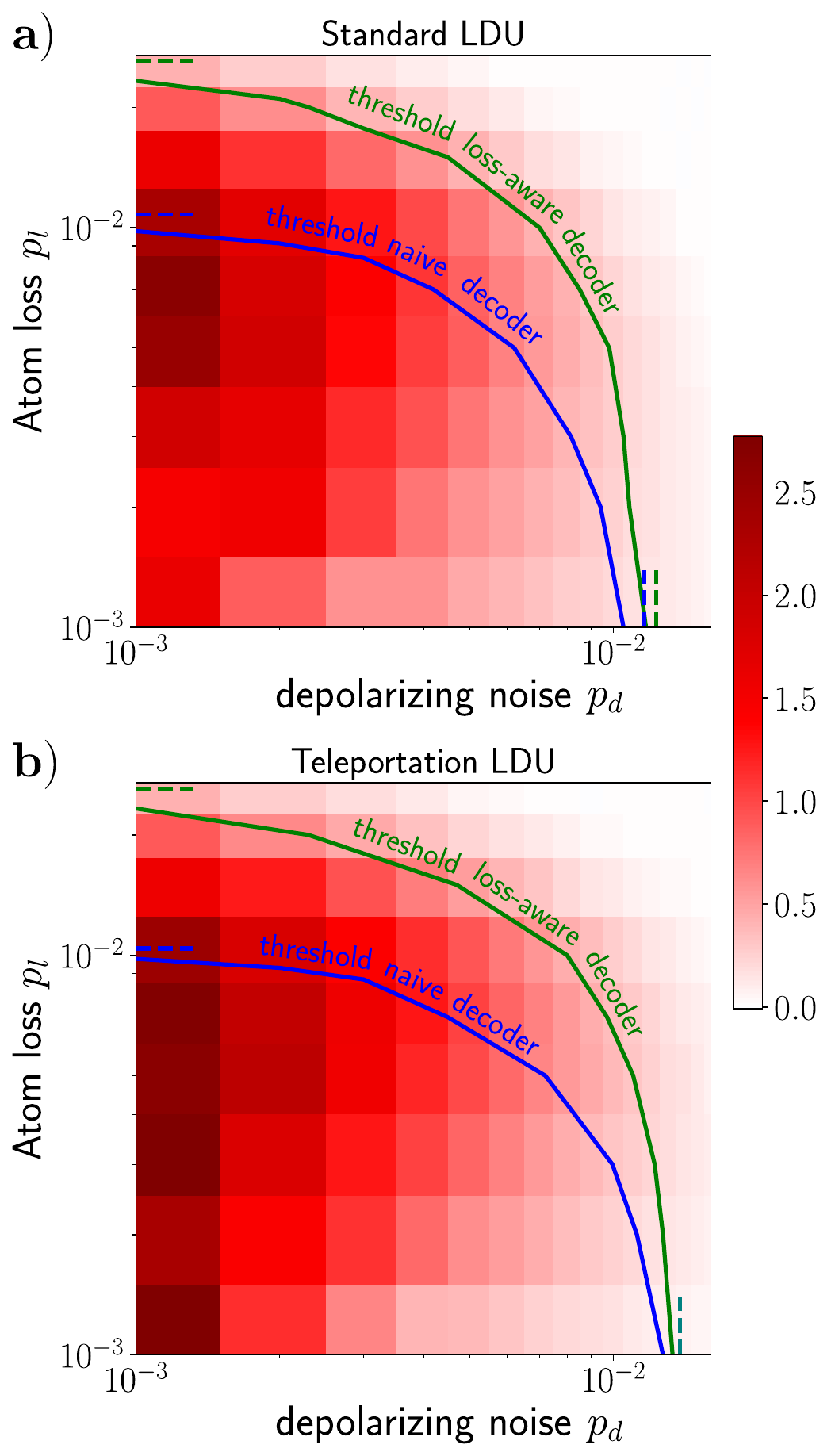}
    \caption{Plot showing the gain of the ``loss-aware'' over the naive decoder as a function of the loss probability $p_l$ and the depolarizing noise error probability $p_d$ for a surface code of distance $d=11$ for ${\bf a)}$ the standard LDU protocol and ${\bf b)}$ the teleportation LDU protocol.}
    \label{fig:gain_decoder}
\end{figure}

We plot the gain $G_\text{LDU}^\text{decoder}$ for both LDU protocols in Fig.~\ref{fig:gain_decoder} in the case of a surface code of distance $d=11$. One important outcome of this study is that the ``loss-aware'' decoder significantly outperforms the naive decoder for both LDU protocols, particularly below the threshold and when $p_d\lesssim p_l$. The maximum gain reaches comparable values of up to $2.8$ for the teleportation LDU protocol and $2.7$ for the standard LDU protocol, for sufficiently small $p_d$. 

In the figure, we further draw the computed threshold lines for both decoders: The data confirm that the ``loss-aware'' decoder has a higher threshold for all values of $p_l$ and $p_d$ for both LDUs. We note that the behavior of the decoders differs slightly for a vanishing loss probability $p_l = 0$. In fact, the noise threshold is identical for both decoders in the teleportation LDU protocol (see overlapping dashed vertical green and blue lines in Fig.~\ref{fig:gain_decoder}{\bf b}), as loss events no longer occur in the simulations. For the standard LDU (Fig.~\ref{fig:gain_decoder}{\bf a}), however, the noise thresholds differ slightly since depolarizing noise can flip measurement outcomes, causing the LDU to incorrectly signal a loss. In that case, the ``loss-aware'' decoder has a slightly higher noise threshold.

In the opposite regime where the loss probability is stronger than the depolarizing noise error probability ($p_l \gg p_d$), the thresholds of the two decoders differ notably: For vanishing depolarizing noise and finite loss probability, the threshold for the naive decoder is found to be approximately $1\%$ (see Fig.~\ref{fig:naive no depo} in Appendix~\ref{ap:naive error}) compared to $2.6\%$ for the ``loss-aware'' decoder .

\section{Simulation}
\label{sec:sim_method}

 The simulation of the surface code with LDUs  is conducted using Stim~\cite{Gidney2021}. First, for each shot of the simulation, a sample of lost atoms is randomly drawn from a probability distribution described in this section. Based on this sample, a new circuit is recompiled by removing the appropriate gates. Finally, the modified circuit is simulated for one run using Stim's simulator. For each pair $(p_l,p_d)$ and for various code sizes from $d=3$ to $d=11$, $10^5$ shots are performed to determine the logical error probability ($10^6$ shots if the logical error probability is $\lesssim 10^{-5}$).

    \subsection{Loss probabilities}
    \label{sec:loss_proba}

 We generate the sample of lost atoms by randomly determining, for each atom in each cycle, whether the atom is present or, if absent, the position at which it was lost, drawn from a suitable distribution $\{p_{l,i}\}_{i=0}^{n^{\text{tot.}}_{\text{CZ}}}$. Here, $p_{l,i}$ is the probability to lose the atom at the $i$-th CZ gate within a round for $i>0$. $p_{l,0}=1-\sum_{i=1}^{n^{\text{tot.}}_{\text{CZ}}}p_{l,i}$  depicts the probability that the atom is not lost. $n^{\text{tot.}}_\text{CZ}$  represents the total number of CZ gates experienced by the atom in a round.

 The probability of an atom being lost in a cycle depends on the number of CZ gates applied to it.  Data atoms in the bulk (resp. on the edges, in the corners) are coupled through CZ gates to four (resp. three, two) ancilla qubits. Additionally, each of these data atoms is coupled to the LDU's ancilla, with two CZ gates in the case of the standard LDU, or one CZ gate in the teleportation-based LDU. In the final round, all qubits are measured, and no LDU is required to infer the presence of atoms. Ancilla atoms used for stabilizer measurements are either located in the bulk, coupled to four data qubits, or on the edges, coupled to two data atoms.  They do not employ any LDU.

\subsubsection{Loss probabilities for stabilizer operations}

First, we compute the probability to lose an atom during the stabilizer operations.  In our simulation, we first apply the Z stabilizers in parallel followed by the X stabilizers.  The loss probability of each individual CZ gate is denoted as $p_l$. For each atom, the probability to be lost at the $i$-th CZ gate in a round is given by Eq.~\eqref{eq:usual_loss}.

As the operations seen by the stabilizer ancilla atoms only involve stabilizer operations, the loss distribution $\{p_{l,i}\}_{i=0}^{n^{\text{tot.}}_{\text{CZ}}}$ follows Eq.~\eqref{eq:usual_loss}. Here $n^{\text{tot.}}_{\text{CZ}}=3$ or $4$ depending on whether the ancilla atom is located on the edge or in the bulk, respectively.

\subsubsection{Simulation of the teleportation-based LDU}
 For the sake of efficiency, LDUs are never directly simulated  in Stim as their effect is trivial in the absence of errors. However, their impact can be emulated in the presence of loss or Pauli noise. For example, if a data atom is lost, it is reset to $\ket{0}$ at the end of the cycle to mimic the reloading of a new atom. We also account for losses and noise on the CZ gates during the LDU (see Sec.~\ref{sec:sim_depo}).

\begin{figure*}[t]
    \includegraphics[width=\linewidth]{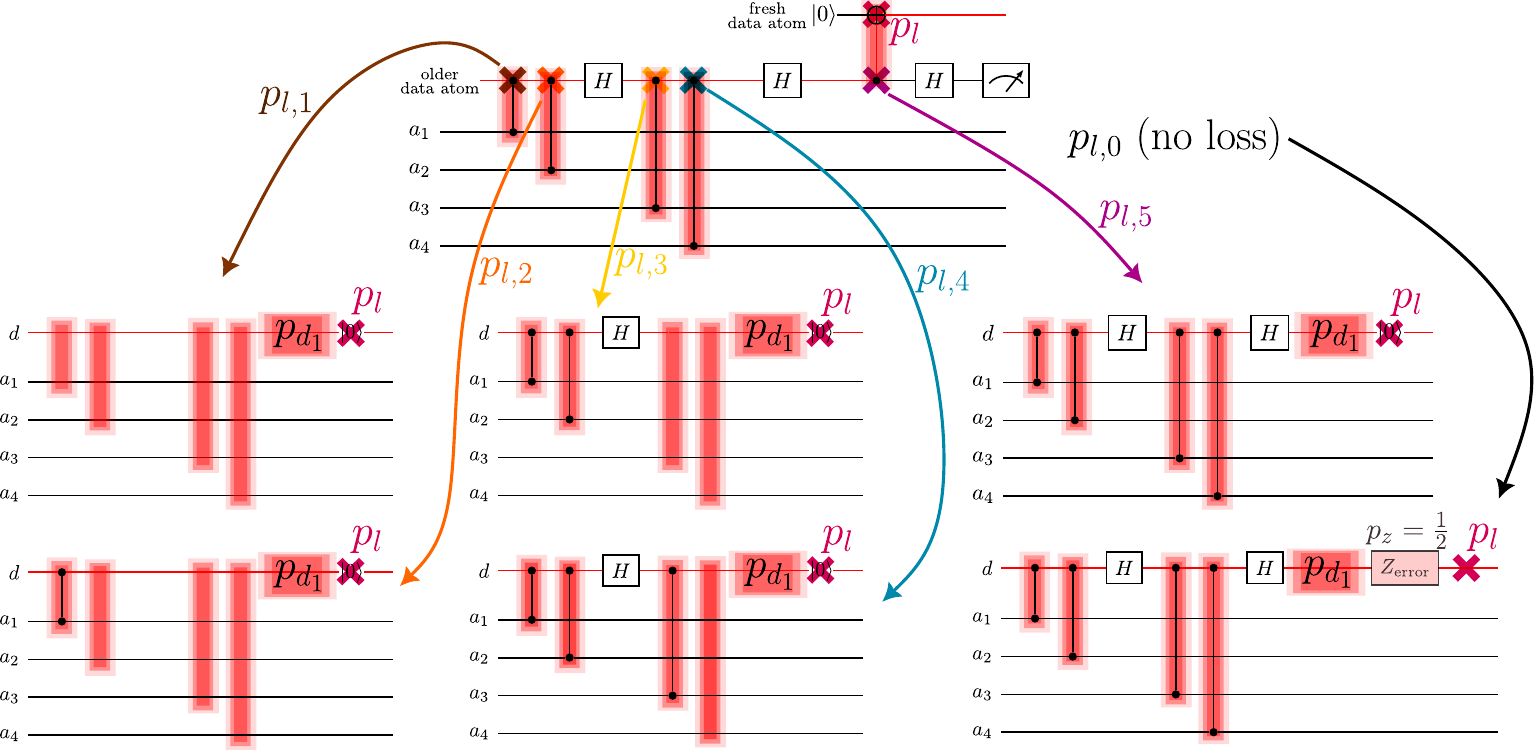}
    \caption{Operations seen by a bulk data atom at each cycle when the surface code is equipped with the teleportation LDU. The original circuit with the teleportation LDU (top circuit) is never simulated. In the absence of atom loss (bottom right circuit), the teleported state exhibit a Z error $50\%$ of the time. The data atom being simulated throughout the round is represented by the red wire, which transitions from the older data atom to the newly loaded one during the teleportation protocol. To simulate atom loss, gates are erased in the circuit.  Crosses indicate the potential loss locations in the original circuit. The arrows show the effective circuits being simulated on Stim along with their probabilities $p_{l,i}=p_l(1-p_l)^{i-1}$ (see Eq.~\eqref{eq:usual_loss}). The fresh atom can be lost during the LDU with probability $p_l$ (pink crosses), resulting the data atom to not be reloaded in the following cycle (i.e. the next round will be similar to the middle left effective circuit). Red boxes represent the depolarizing noise channel associated with CZ gates. The resulting 1-qubit effective noise channel stemming from the noisy teleportation LDU has an error probability $p_{d_1}^{\text{tel. LDU}}$ (see Eq.~\eqref{eq:effective error rate teleportation LDU}), which we denote simply as $p_{d_1}$ in the figure to avoid visual clutter. }
    \label{fig:eff circuit tel LDU}
\end{figure*}

 Consequently, in the quantum circuit representation of the teleportation LDU protocol, the wire of the data atom  represents the "old" data atom {\it before} the teleportation protocol and the "fresh" atom {\it after} the teleportation protocol (see red wire of the upper circuit in Fig.~\ref{fig:eff circuit tel LDU}).
 
When supplementing the surface code with the teleportation LDU protocol, each data atom undergoes an additional CZ gate due to the teleportation LDU, on top of the usual stabilizer operations. The distribution $\{p_{l,i}\}_{i=0}^{n^{\text{tot.}}_{\text{CZ}}}$ remains as defined by Eq.~\eqref{eq:usual_loss}, with $n^{\text{tot.}}_\text{CZ}=3$ or $4$ or $5$ depending on whether the atom is located in the corner, on the edge or in the bulk, respectively.

Additionally, the freshly loaded atom can be lost during the LDU with a probability $p_l$, causing it to remain absent in the subsequent cycle. This is accounted for in our simulation by erasing the gates involving that atom in the following round, regardless of the result of the random draw for that atom in the cycle.

Fig.~\ref{fig:eff circuit tel LDU} illustrates the different effective circuits simulated along with their probabilities.
\subsubsection{Simulation of the standard LDU}
 For the CZ gates involved in the standard LDU, the probability of loss is slightly higher because the LDU can be reapplied if the ancilla is lost, leading the data qubit to experience, on average, more CZ gates.  Instead of directly simulating this conditional reapplication, we mimic this procedure by increasing the loss probability of the associated CZ gates. The probability of losing both the data qubit and the LDU's ancilla is given by:
    \begin{align}
        p_{l,\text{LDU}0}&=(1-p_l)^{n^{\text{stab.}}_\text{CZ}+2}\sum_{r=0}^\infty(1-(1-p_l)^2)^r p_l\nonumber\\ &\times (2-p_l)\sum_{i=0}^{r-1}(1-p_l)^{2i}
    \end{align} where $n^{\text{stab.}}_\text{CZ}$ is the number of CZ gates used in the standard stabilizer measurements, the index $r$ indicates  the number of LDUs applied due to the loss of the ancilla along with its probability $(1-(1-p)^2)^r$. The factor $(1-p_l)^{n^{\text{stab.}}_\text{CZ}+2}$ accounts for the probability that the data atom is not lost during the previous $n_\text{CZ}^{\text{stab.}}$ CZ gates used for stabilizer measurements, and that the LDU's ancilla  is not lost during the final LDU. The last term $p(2-p)\sum_{i=0}^{r-1}(1-p)^{2i}$ expresses the probability to lose the data qubit during the application of the various LDUs. 

 The probability to lose the data atom on the first CZ gate of the last LDU is given by
    \begin{align}
    p_{l,\text{LDU}1}=(1-p_l)^{n^{\text{stab.}}_\text{CZ}+2}\sum_{r=0}^\infty&(1-(1-p_l)^2)^r\nonumber\\
    &\times p_l(1-p_l)^{2r}.
    \end{align}

 Together, $p_{l,\text{LDU}0}$ and $p_{l,\text{LDU}1}$ lead to the same effective circuit where the two CZ gates of the standard LDU are suppressed because of the depletion of the data atom. In any case, this loss will be detected by the final LDU (assuming no depolarizing errors) and will trigger the reloading of a new atom in the subsequent cycle. The combined probabilities can be summed, yielding
    \begin{align}
p_{l,n^{\text{stab.}}_\text{CZ}+1}&=p_{l,\text{LDU}0}+p_{l,\text{LDU},1}\\
    &= \frac{p_l(1-p_l)^{n^{\text{stab.}}_\text{CZ}}(2-p_l-(1-p_l)^3)}{1-(1-p_l)^2+(1-p_l)^4}.
    \label{eq:loss stand. LDU 1}
    \end{align}
 Similarly, the probability of losing the data qubit during the second CZ gate of the final standard LDU is given by
 \begin{align}
    p_{l,n^{\text{stab.}}_\text{CZ}+2}=&  (1-p_l)^{n^{\text{stab.}}_\text{CZ}+2}\sum_{r=0}^\infty(1-(1-p_l)^2)^r \nonumber\\& \times p_l(1-p_l)^{2r+1}\\
    =&\frac{p_l(1-p_l)^{n^{\text{stab.}}_\text{CZ}+3}}{1-(1-p_l)^2+(1-p_l)^4}.
     \label{eq:loss stand. LDU 2}
    \end{align}

\begin{figure*}[t]
    \includegraphics[width=\linewidth]{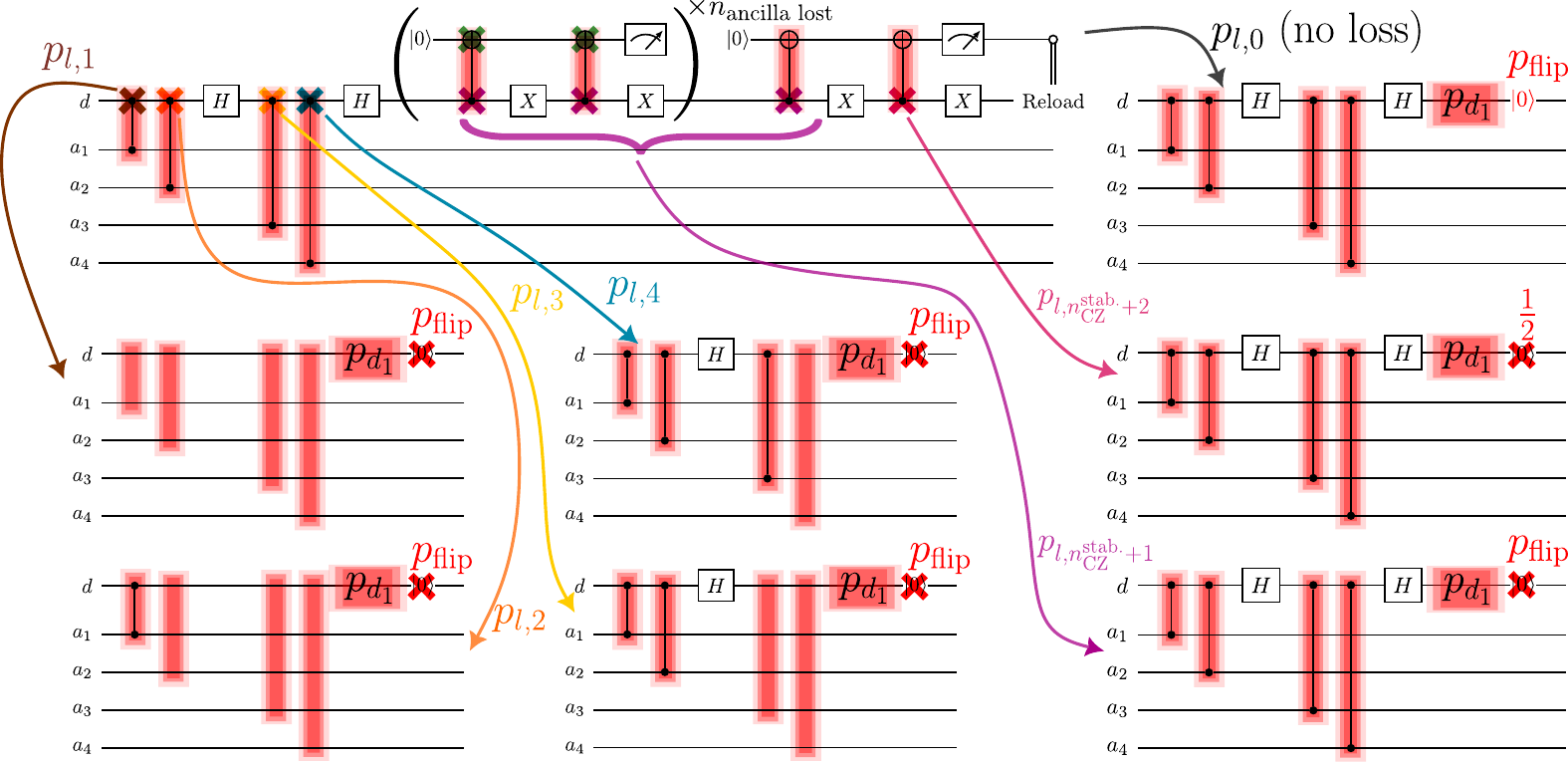}
    \caption{Operations seen by a bulk data atom at each cycle when the surface code is equipped with the standard LDU. The original circuit with the standard LDU (top left circuit) is never simulated. In the absence of atom loss (top right circuit), the action of the standard LDU is trivial.  To simulate atom loss, gates are erased in the circuit.  Crosses indicate the potential loss locations in the original circuit. The arrows show the effective circuits being simulated on Stim along with their probabilities $p_{l,i}$ (see Eq.~\eqref{eq:usual_loss} for $0<i\leq 4 $ and Eq.~\eqref{eq:loss stand. LDU 1} and Eq.~\eqref{eq:loss stand. LDU 2} for $i=n^{\text{stab.}}_\text{CZ}+1,n^{\text{stab.}}_\text{CZ}+2$). $n_\text{ancilla lost}$ is the number of time the LDU's ancilla is lost (its potential loss locations are indicated by green crosses) during the standard LDU. This loss can be detected and another LDU is reapplied in that case. Red boxes represents the depolarizing noise channel associated with CZ gates. The resulting 1-qubit effective noise channel stemming from the noisy standard LDU has an error probability $p_{d_1}^\text{stand. LDU}$ (see Eqs.~\eqref{eq:fidelity standard LDU},~\eqref{eq:effective error rate standard LDU}) , which we denote simply as $p_{d_1}$ in the figure to avoid visual clutter. Noise in the standard LDU also induces a flip with probability $p_\text{flip}$ (see Eq.~\eqref{eq:proba flip}) in the measurement outcome causing the loss data atom not being reloaded (red crosses) or reloading an new atom while still being present (red $\ket{0}$ on the top right circuit).}
    \label{fig:eff circuit stand LDU}
\end{figure*}

The probability that the standard LDU detects the loss on the last CZ gate is $50\%$. Consequently, with probability  $p_{l,n^{\text{stab.}}_\text{CZ}+2}/2$, the data qubit may remain lost in the subsequent cycle (see middle right circuit in Fig.~\ref{fig:eff circuit stand LDU}). \\

In the standard LDU protocol, the loss probability distribution $\{p_{l,i}\}_{i=0}^{n^{\text{tot.}}_{\text{CZ}}}$ is therefore given by Eq.~\eqref{eq:usual_loss} for $0<i\leq n^{\text{stab.}}_\text{CZ} $, by Eq.~\eqref{eq:loss stand. LDU 1} for $i= n^{\text{stab.}}_\text{CZ}+1$ and Eq.~\eqref{eq:loss stand. LDU 2}
 for $i= n^{\text{stab.}}_\text{CZ}+2=n^{\text{tot.}}_{\text{CZ}}$. 
Fig.~\ref{fig:eff circuit stand LDU} illustrates the different effective circuits simulated on Stim along with their probability when the surface code is supplemented with the standard LDU.

\subsection{Depolarizing noise}
    \label{sec:sim_depo}
     In this section, we address the simulation of the depolarizing noise channel. Each CZ gate employed in the stabilizer measurements is dressed with a two-qubit depolarizing noise channel with error probability $p_d$ implemented using the standard method available in Stim. 
     
     When CZ gates are removed due to atom loss, we retain their associated noise channel since pulses are always applied, even in the absence of atoms. The effect of the noise channel on the lost atom is trivial because it does not entangle with other qubits until a new atom (modelled by a reset operation in the effective circuits, see Fig.~\ref{fig:eff circuit tel LDU},~\ref{fig:eff circuit stand LDU}) is reloaded.
     
    For CZ gates in LDUs, a careful analysis of the impact of their noise is required, as LDUs are not directly simulated using Stim. Their effect is twofold: First, they induce an effective 1-qubit depolarizing channel on the data atom they are coupled to; Second, the presence of depolarizing noise may flip the measurement outcome of the LDU.
    
    \subsubsection{Effective 1-qubit depolarizing channel}
    A 2-qubit depolarizing noise channel can be written as
    \begin{align}
       \mathcal{E}(\rho)&=(1-p_d)\rho+\frac{p_d}{15}\sum_{\substack{i,j=0\\(i,j)\neq (0,0)}}^3 \sigma_i\otimes\sigma_j.\rho.\sigma_i\otimes\sigma_j\label{eq:depo1}\\
        &=(1-\frac{16 p_d}{15})\rho+\frac{16 p_d}{15}\frac{\mathbb{I}_4}{4},
        \label{eq:depo2}
    \end{align}
    where $\sigma_k$ with  $k=1,2,3$ denote the $X,Y,Z$  Pauli matrices, respectively, and $\sigma_0=\mathbb{I}_2$ is the $2\times2$ identity matrix.
     To calculate the resulting error probability $p_{d_1}^{\text{LDU}}$ for the effective 1-qubit depolarizing channel, a convenient approach is to change the representation of the error channel. By using the Pauli transfer matrices (PTM) formalism~\cite{vandenBerg2023a}, the density matrix is promoted as a vector in a new Hilbert space and noise channels become matrices. The impact of the noise channel at the end of the LDU circuit can be evaluated by computing the matrix product $N(C).C^{-1}$ where $C$ (resp.$N(C)$) is the PTM representation of the noiseless (resp. noisy) LDU circuit. To obtain the final noise channel, we switch back to the standard representation of the error channel (e.g. as in Eq. \eqref{eq:depo1}, or Kraus representation) and trace over the ancilla or older data atom degrees of freedom yielding the effective 1-qubit noise channel.
     
     In the particular case of a depolarizing noise channel, the PTM representation is proportional to the identity matrix, scaled by its fidelity. The term $N(C).C^{-1}$ becomes equivalent to the computation of the total fidelity of the LDU circuit, denoted as $f_d^{\text{LDU}}$. If all depolarizing noise channels have similar error probabilities, the latter is expressed as $f_d^{\text{LDU}}=f^{n_\text{noise}}$,  where $f$ is the fidelity of a single noise channel and $n_\text{noise}$ is the number of depolarizing noise channels in the LDU circuit. Finally, by tracing over the ancilla degrees of freedom in the standard LDU, or over the older data atom degrees of freedom in the teleportation LDU, we derive the following formula for the error probability of the effective 1-qubit noise channel
     \begin{align}
        p_{d_1}^{\text{LDU}}=\frac{3}{4}(1-f_{d}^{\text{LDU}})\quad .
        \label{eq:effective error rate}
    \end{align}

      For the teleportation-based LDU protocol, as there is only one CZ gate, the fidelity is directly the fidelity of the 2-qubit noise channel
    \begin{align}
        f_{d}^{\text{tel. LDU}}=1-\frac{16p_d}{15}\quad .
    \end{align}
    Using Eq.~\eqref{eq:effective error rate}, the resulting error probability of the effective 1-qubit depolarizing channel is
    \begin{align}
        p^{\text{tel. LDU}}_{d_1}=\frac{4p_d}{5}\quad .
        \label{eq:effective error rate teleportation LDU}
    \end{align}
      For the standard LDU protocol, since we can reapply the LDU if the ancilla has been lost, computing the fidelity requires summing over the different number of LDUs applied due to these losses along with their probability. The total circuit fidelity reads
    \begin{align}
        f_{d}^{\text{stand. LDU}}&=\sum_{r=0}^\infty(1-(1-p_l)^2)^r\left(1-\frac{16p_d}{15}\right)^{2r+2\nonumber}\\
        &\times (1-p_l)^2\\
                &=\frac{\left(1-\frac{16p_d}{15}\right)^{2}(1-p_l)^2}{1-(1-(1-p_l)^2)(1-\frac{16p_d}{15})^2}\quad .
            \label{eq:fidelity standard LDU}
    \end{align}
    As above, using Eq.~\eqref{eq:effective error rate}, we implement on Stim a 1-qubit depolarizing channel of error probability 
    \begin{align}
        p_{d_1}^{\text{stand. LDU}}=\frac{3}{4} \frac{1-(1-\frac{16p_d}{15})^2}{1-(1-(1-p_l)^2)(1-\frac{16p_d}{15})^2}\quad.
        \label{eq:effective error rate standard LDU}
    \end{align}
    \subsubsection{Measurement flips}
     Depolarizing noise can also cause measurement errors through its Pauli error terms, which anticommute with the measurement basis.     
     In the teleportation-based LDU protocol, the older data atom is measured in the X-basis (achieved by employing an Hadamard gate followed by a Z-basis measurement). As a result, Y and Z errors from the depolarizing noise channel on the older atom will lead to incorrect measurement outcomes, resulting in a Z error on the new atom during the decoding process. Equation~\eqref{eq:depo1} then becomes
     \begin{align}
         \mathcal{E}(\rho)&=(1-p_d)\rho \nonumber\\ &+\frac{p_d}{15} \sum_{\substack{i=0\\(i,j)\neq (0,0)}}^3\sum_{j=0}^1\sigma_i\otimes\sigma_j.\rho.\sigma_i\otimes\sigma_j\nonumber\\
         &+ \frac{p_d}{15}\sum_{i=0}^3 \sum_{j=2}^3 (Z.\sigma_i)\otimes\sigma_j.\rho.(\sigma_i.Z)\otimes\sigma_j
     \end{align}
 where the first index $i$ represents the fresh atom and the second index $j$ refers to the older atom.  The resulting noise channel remains depolarizing with the same error probability due to the isotropy of the noise, making the impact of this incorrect Z error trivial in that case.

In the case of the standard LDU, two effects are expected. The LDU may indicate the loss of an atom while still being present. This leads to the reloading of a new atom initialized in state $\ket{0}$. We assume that before reloading, we allow any remaining atom to leave, preventing collisions with the newly loaded atom. On the other hand, there is a possibility that the LDU fails to detect the actual loss of an atom, resulting on its absence potentially over many rounds. For a depolarizing noise channel, this non detection occurs with the same probability as the false loss detection. The probability that the measurement result will be flipped due to the depolarizing noise is given by 
    \begin{align}
        p_{\text{flip}}&=\frac{1-(1-\frac{16 p_d}{15})^2}{2}\quad .
        \label{eq:proba flip}
    \end{align}

\section{Loss with maximally biased Z error}
\label{sec:real_loss}
\par In this final section, we modify the loss model and evaluate the resulting performance of the surface code. Experimentally, atomic loss caused by the Rydberg anti-trapping effect (or leakage outside the computational state) can occur at any time during the pulse application of the CZ gate. When this happens, the remaining atom is projected into the $\ket{1}$ state (due to the Rydberg blockade). The remainder of the pulse subsequently re-excites the remaining atom, leaving it in the Rydberg state, which then decays into a state outside the computational subspace with high probability~\cite{Wu2022}. In practice, the atom may decay outside the computational subspace with high probability. As already mentioned in Sec.~\ref{sec:loss}, we leave this correlated loss mechanism as a future direction for work and focus here only on independent loss. Another possibility is to decay back to the qubit subspace. Here, additionally, we assume that the decay to the $\ket{0}$ state is forbidden and the atom can only decay back to the $\ket{1}$ state. This can be realized in experiments by carefully encoding the qubit states into well-chosen atomic levels~\cite{Cong2022}.
 Using RC, this noise channel can be effectively transformed into a maximally (i.e. error probability of 0.5) biased Z-error channel~\cite{Sahay2023} . Therefore, our revised loss model is the following: whenever an atom is lost during a given CZ gate, the neighboring remaining atom will experience a Z error of probability $0.5$.

In this section, we focus on the teleportation LDU as it shows the best performance for the simple loss model.  Moreover, we anticipate the standard LDU to perform even worse than before compared to the teleportation LDU under this new loss model. This is because the conditional reapplication of the standard LDU, based on the loss detection of its ancilla increases the  probability of making a Z error on the adjacent data qubit.

\begin{figure}
    \centering
    \includegraphics[width=\linewidth]{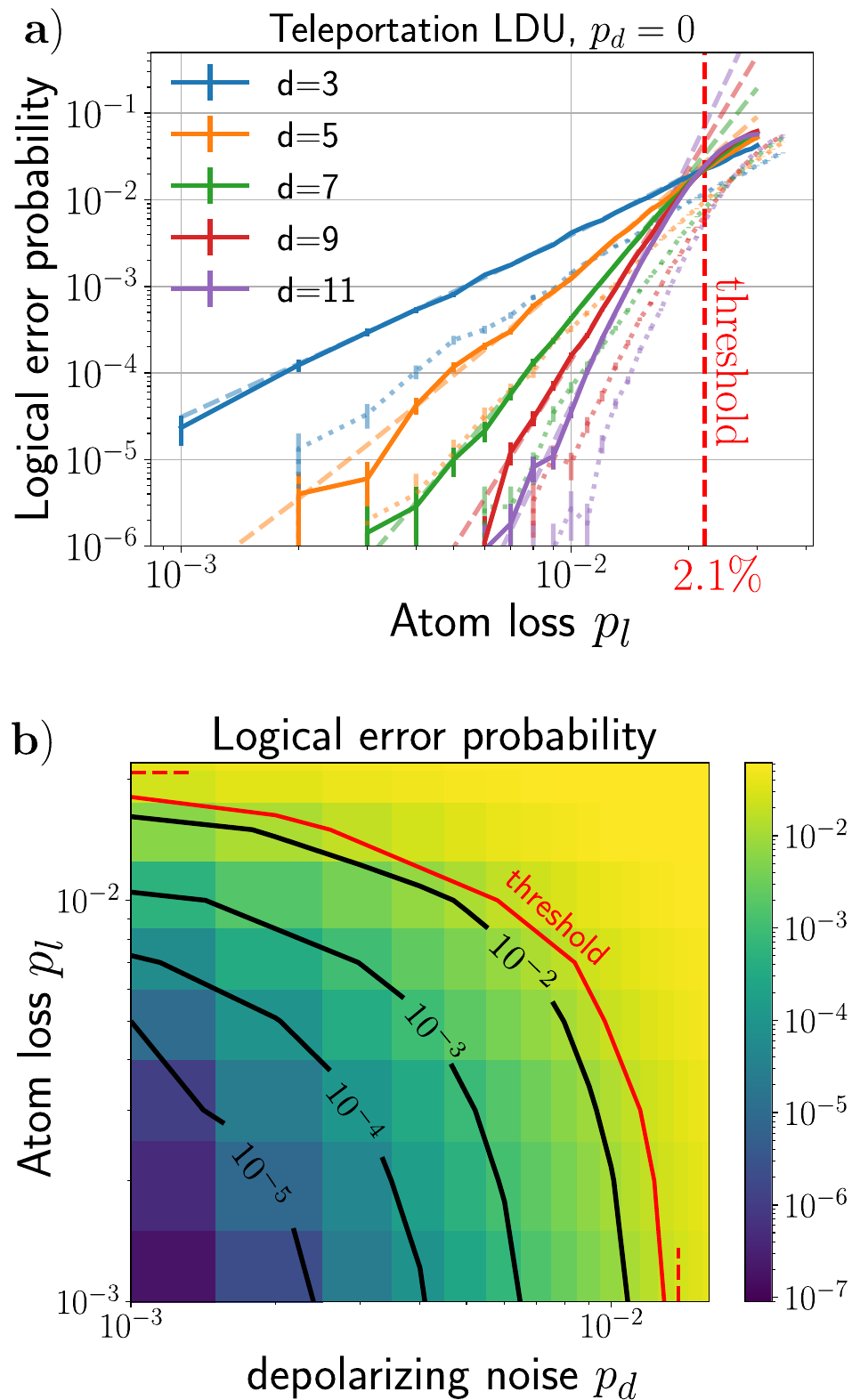}
    \caption{Performance analysis of the teleportation LDU scheme under the revised loss model. {\bf a)} The logical error probability, normalized by the number of rounds for code distances $d=3,5,7,9,11$ as a function of the atom loss probability $p_l$ and at vanishing depolarizing noise $p_d=0$ (solid lines). The red vertical dashed line indicates the error threshold at $2.1\%$. Other color dashed lines show fitted curves using Eq.~\eqref{eq:fit realistic loss model}. For comparison, dotted lines  represent results from for the simpler loss model studied in the rest of the article (see Fig.~\ref{fig:no depo}{\bf b}). {\bf b)} Two dimensional color map of the logical error probability, normalized by the number of rounds for distance $d=11$ as a function of the loss probability $p_l$ and the depolarizing error probability $p_d$. The red solid line marks the error threshold while black solid lines show curves of constant logical error probability. The horizontal and vertical red dashed lines indicate the threshold at vanishing depolarizing noise and loss probability, respectively. Statistics were gathered using $10^5$ shots, except for the region below the $10^{-5}$ logical error probability contour in {\bf b)} where $10^6$ shots were employed.}
    \label{fig:revised loss model}
\end{figure}

Fig.~\ref{fig:revised loss model}  presents the results obtained using the revised loss model.  At vanishing depolarizing noise $p_d=0$ (see Fig.~\ref{fig:revised loss model}{\bf a}), we observe a loss threshold of $2.1\%$, which is lower than the $2.6\%$ threshold found with the simpler loss model. For loss rates below $1\%$, the logical error probability per round is approximately one order of magnitude higher than in the simplified loss model (shown by dotted lines in Fig.~\ref{fig:revised loss model}{\bf a}). Since qubit loss affects neighboring qubits in this model, we expect the power law scaling exponent to be $(d+1)/2$, compared to an exponent of $d$ in the simpler loss model. We fit our numerical results using the following function:
\begin{align}
    \sum_{k= \frac{d+1}{2}}^{d} a_{k}(d) p_l^k
    \label{eq:fit realistic loss model}
\end{align}
 where $a_k(d)$ are fitting parameters that represents the number of weight-k physical errors that lead to a logical error for a code distance $d$.
For the experimentally relevant parameter range ($0.001\leq p_l \leq 0.01$), our fitting analysis reveals an unexpected result: the dominant terms in Eq.~\eqref{eq:fit realistic loss model} correspond to exponent close to $d$ (e.g. $k=d-2,d-1$ for $d=5,7,9,11$). The list of values for the coefficients $a_k(d)$ are summarized in Table~\ref{tab:coef}. Values below $1$ have been rounded to $0$. While these values are not strictly zero, the number of shots is too small to accurately assess them.

\begin{table*}
\centering
\begin{tabular}{|c||c|c|c|c|c|c|}
\hline
$d$& \multicolumn{6}{|c|}{Coefficients $a_k(d)$} \\
\hline
3&\makecell{{\footnotesize \it k=2} \\ 30} &   \makecell{{\footnotesize \it k=3} \\ 935}&\tikzmark{start1}\tikzmark{end1}&&&\\
\hline 5&\makecell{{\footnotesize \it k=3} \\231}&\makecell{{\footnotesize \it k=4}\\$10^{5}$}&\makecell{{\footnotesize \it k=5}\\0}& \tikzmark{start2}\tikzmark{end2}&& \\
\hline 7&\makecell{{\footnotesize \it k=4}\\0}&\makecell{{\footnotesize \it k=5}\\$2.3\times 10^{6}$} &\makecell{{\footnotesize \it k=6} \\ $1.9\times 10^{8}$} & \makecell{{\footnotesize \it k=7}\\0}& \tikzmark{start3}\tikzmark{end3}&     \\
\hline 9 &\makecell{{\footnotesize \it k=5}\\0 }&\makecell{{\footnotesize \it k=6}\\5}&\makecell{{\footnotesize \it k=7}\\$1.1\times 10^{10}$}& \makecell{{\footnotesize \it k=8}\\$3.5\times 10^{11}$}&\makecell{{\footnotesize \it k=9}\\$3.7\times 10^{8}$}& \tikzmark{start4}\tikzmark{end4}\\ 
\hline
11 & \makecell{{\footnotesize \it k=6}\\ 140} & \makecell{{\footnotesize \it k=7}\\$2.2\times 10^{4}$}&\makecell{{\footnotesize \it k=8} \\$2.6\times 10^{4}$} &\makecell{{\footnotesize \it k=9}\\$3.6\times 10^{13}$} &\makecell{{\footnotesize \it k=10} \\$2.1\times 10^{6}$}& \makecell{{\footnotesize \it k=11} \\ $5.7\times 10^{16}$}
\\ \hline

\end{tabular}

\caption{Coefficients $a_k(d)$ obtained by fitting the simulation results from the realistic loss model at vanishing depolarizing noise $p_d=0$ (see Fig.~\ref{fig:revised loss model}{\bf a}) using Eq.~\eqref{eq:fit realistic loss model}.}
    \label{tab:coef}

\begin{tikzpicture}[overlay, remember picture]
  % Hatching for the first marked cell
  \fill[pattern=north east lines, pattern color=black] 
    ($(pic cs:start1)+(-1.05cm,-0.38cm)$) rectangle 
    ($(pic cs:end1)+(7.2cm,0.57cm)$);

\fill[pattern=north east lines, pattern color=black] 
    ($(pic cs:start2)+(-1.05cm,-0.38cm)$) rectangle 
    ($(pic cs:end2)+(5.1cm,0.57cm)$);

\fill[pattern=north east lines, pattern color=black] 
    ($(pic cs:start3)+(-0.97cm,-0.38cm)$) rectangle 
    ($(pic cs:end3)+(3.06cm,0.57cm)$);
\fill[pattern=north east lines, pattern color=black] 
    ($(pic cs:start4)+(-1.04cm,-0.38cm)$) rectangle 
    ($(pic cs:end4)+(1.04cm,0.57cm)$);    
\end{tikzpicture}
\end{table*}

When both  atom loss and depolarizing noise are present (see Fig.~\ref{fig:revised loss model}{\bf b}), the logical error probability is higher and the error threshold is lower than in the simplified loss model. Nevertheless, both models exhibit qualitatively similar behavior.

\section{Conclusion}
 In conclusion, this study demonstrates that the integration of LDUs with the surface code enables efficient QEC in the presence of both atom loss and depolarizing noise, provided losses and depolarizing errors remain below the threshold.
 Specifically, for a surface code with distance $d$ and no depolarizing noise, it is possible to correct up to $d-1$ errors under the first loss model, where the loss of an atom does not affect the neighboring qubit. This behavior is similar to an erasure channel, although only partial knowledge of the loss location is available through the LDUs. When the loss impacts neighboring atoms, the scaling law is slightly affected, with the logical error probability scaling approximately with a power law of exponent $\sim d -2$. For both loss models, the system is highly resilient to atom loss.

 When both noise sources—atom loss and depolarizing noise—are present, the scaling of the logical error probability is primarily driven by depolarizing errors. In such cases, the system can correct up to $\frac{d-1}{2}$
errors, but both types of errors contribute to the overall logical failure. By designing a decoder able to leverage loss locations, we significantly enhance the system’s performance compared to a naive decoding approach. This improvement allows for a reduction in the logical error probability by nearly three orders of magnitude for a code distance of $d=11$, particularly when the loss probability exceeds the depolarizing noise error probability.

 Furthermore, our results show that the teleportation-based LDU protocol exhibits a smaller logical error probability compared to the standard LDU. This improvement is due to the fact that the teleportation-based protocol requires fewer controlled-Z (CZ) gates, reducing the overall probability of errors arising during gate operations.
 Eventually, the combination of LDUs and surface code provides a promising pathway to achieving robust, fault-tolerant quantum computation in neutral atom systems.\\

 Several research directions could be explored to further refine and improve the error models and the error correction protocol for neutral atom quantum processors. One direction is to investigate more realistic noise models by including imperfect measurements, noisy single-qubit gates or customized CZ gate Pauli noise channels tailored specifically for neutral atoms, providing a more accurate representation of the noise environment in these systems~\cite{Jandura2024}. Additionally, recent studies have shown that leakage out of the computational subspace can be converted to erasure channel~\cite{Ma2023,Wu2022,Scholl2023,Yu2024}, which enhance the overall error correction capability of the system. 
 
 Another direction is to complexify the loss model by distinguishing losses due to heating and collisions, typically occurring between gate applications, from those during CZ gates caused by anti-trapping effects. The former type of loss should be proportional to the time spent between two consecutive CZ gates while the latter may impact the remaining atom involved in the CZ gate by enhancing leakage mechanism. Two directions can be explored: (i) assume the ability to repump the leaked atom back into the computational state, such as considered in Sec.~\ref{sec:real_loss} where the remaining atom experienced a maximally biased Z error after RC or (ii) consider leakage-to-loss conversion schemes which may introduce correlated losses between atoms involved in the same CZ gate (see Supplementary Information of~\cite{Wu2022}).
 %; however due to the Rydberg blockade effect, it is unlikely that both atoms would be simultaneously lost.
 Moreover, an imperfect reloading scheme could be considered, where atom loss might occur during shuttling.

 Experimentally, the time required to perform LDUs could  limit the QEC protocol's performance. 
 It would be interesting to explore the resilience of the surface code augmented by LDUs  when the frequency of LDU applications is diluted (e.g., by performing one LDU every two or more rounds of quantum error correction). %
 
The present analysis can be readily extended to investigate different QEC protocols for neutral atom quantum computing such as the recently introduced high-rate quantum LDPC codes~\cite{Xu2024,Hong2024,Chandra2024,Bravyi2024,Pecorari2025,Pecorari2025a} 

\textit{Note - Several recent studies have also investigated atom loss across different QEC protocols~\cite{Yu2024,Ziad2024,Baranes2025,YU2025}.}

\section*{Acknowledgements}
S. J. gratefully acknowledges discussions with Mikhail Lukin. H. P. and G. P. would like to express their gratitude for insightful discussions with Jeff Thompson. This work was founded by the European Union’s Horizon Europe program
HORIZON-CL4-2021-DIGITAL-EMERGING-01-30 via the project 101070144 (EuRyQa) and by the Horizon 2020 program under
the Marie Sklodowska-Curie project 955479 (MOQS) and by the French National Research
Agency under the Investments of the Future Program
project ANR-21-ESRE-0032 (aQCess).

\section*{Data and code availability}
Data and the code used to generate detector error models, simulate quantum circuits, compute the logical error rates and analyze the results are available on \href{https://zenodo.org/records/14865271}{Zenodo open repository}.

\bibliography{QEC_initials}
\bibliographystyle{quantum}
\newpage
\appendix

\section{Conditional probabilities used in the ``loss-aware'' decoder}
\label{ap:decoding}
 In this Appendix, we explicitly compute the conditional probabilities of the ``loss-aware'' decoder presented in Sec.~\ref{sec:decoder} based on Eq.~\eqref{eq:cond_proba}. 

 Conditioned on the detection of a loss in a given cycle, it exists multiple potential loss locations. The number of potential loss locations differs depending on the type of atom (data or ancilla qubits).\\

  For ancilla qubits, no LDU is required as the measurement outcome gives the information on the presence of the atom. We also assume the measurement to be perfect. If an ancilla qubit is lost during the round $r$, the potential loss locations correspond to the different CZ gates involved in the stabilizer measurement for that round. Each probability is given by:
\begin{align}
p_{\text{pot. loss},i}=\frac{p_{l,i}}{\sum_{k=1}^{n^\text{stab.}_\text{CZ}}p_{l,k}}    \quad .
\end{align} For the ancilla, $p_{l,i}=p_l(1-p_l)^{i-1}$ corresponds to the probability  described by Eq.~\eqref{eq:usual_loss}.

  For data qubits, LDUs are used to identify cycles of loss. When employing the teleportation-based LDU protocol, the data qubit can be lost either during the usual CZ gates involved in the stabilizer measurements or the CZ gate of the teleportation-based protocol of either the previous round or the current round:
\begin{align}
    p_{\text{pot. loss},i}=\frac{p_{l,i}}{\sum_{k=1}^{n^\text{stab.}_\text{CZ}+2}p_{l,k}}
\end{align} where $p_{l,i}$ is the same as for ancilla qubits.

 When using the standard LDU, the decoding process becomes more complex. In the absence of depolarizing noise in the LDU, if a loss is detected at round $r$ and the same LDU has already detected a loss at round $r-1$, the potential loss locations for the loss at round $r$ are restricted to the CZ gates of the current round. Moreover, the loss on the last CZ gate has also 50$\%$ to not be detected:

 \begin{align}
 \forall 1\leq i\leq n^{\text{stab.}}_\text{CZ}+1&,\nonumber\\
    p_{\text{pot. loss},r,i}^{\text{no depo.}}&=\frac{p_{l,i}}{\sum_{k=1}^{n^{\text{stab.}}_\text{CZ}+1}p_{l,k}+ \frac{p_{l,n^{\text{stab.}}_\text{CZ}+2}}{2}}\\
     p_{\text{pot. loss},r,n^{\text{stab.}}_\text{CZ}+2}^{\text{no depo.}}    &=\frac{p_{l,n^{\text{stab.}}_\text{CZ}+2}/2}{\sum_{k=1}^{n^{\text{stab.}}_\text{CZ}+1}p_{l,k}+ \frac{p_{l,n^{\text{stab.}}_\text{CZ}+2}}{2}}
\end{align}
where $p_{l,n^{\text{stab.}}_\text{CZ}+1}$ and $p_{l,n^{\text{stab.}}_\text{CZ}+2}$ are given by Eqs.~\eqref{eq:loss stand. LDU 1},~\eqref{eq:loss stand. LDU 2} respectively.

 In the case where the LDU of the same atom has not detected any loss at the previous cycle, another potential loss location is the last CZ gate of the LDU from the previous cycle, which has a $50\%$ chance of going undetected:

\begin{align}
    p'^{\text{ no depo.}}_{\text{pot. loss},r,i}=\frac{p_{l,i}}{\sum_{k=1}^{n^{\text{stab.}}_\text{CZ}+2}p_{l,k}},\forall 1\leq i\leq n^{\text{stab.}}_\text{CZ}+1
\end{align}

\begin{align}
    p'^{\text{ no depo.}}_{\text{pot. loss},r,n^{\text{stab.}}_\text{CZ}+2}&=p'^{\text{no depo.}}_{\text{pot. loss},r-1,n^{\text{stab.}}_\text{CZ}+2}\\
    &=\frac{p_{l,n^{\text{stab.}}_\text{CZ}+2}/2}{\sum_{k=1}^{n^{\text{stab.}}_\text{CZ}+2}p_{l,k}}
\end{align}

 When turning on depolarizing noise, LDU may wrongly indicate a loss or on the contrary not detect a lost qubit over multiple cycles. To compute the corresponding probabilities, we denote by $r_l$ the last round where the LDU of the same qubit indicated a loss. The probability that the LDU falsely indicates a loss at round $r$ is:
\begin{align}
    p_{\text{no loss},r,r_l}=\frac{p_{\text{flip}} p_{l,0}^{r-r_l}}{p_{\text{tot.},r,r_l}}
\end{align}
where $p_{\text{flip}}$ is defined in Eq.~\eqref{eq:proba flip}, $p_{\text{tot.},r,r_l}$ is the sum of all the probabilities for the potential loss locations we are currently defining in the case of data atoms employing the standard LDU protocol and in the presence of depolarizing noise.   

The probability that the loss occurred  at the $i$-th CZ gate of the round $r_t$ given the LDU indicates a loss at round $r$ is:
\begin{flalign}   
&\forall r_l<r_t\leq r,\ \forall 1 \leq i \leq n_\text{CZ}^\text{stab.}+1,  \nonumber\\
&p_{\text{pot. loss},r_l,r_t,r,i}=\frac{p_{l,i}p_{\text{flip}}^{r-r_t}(1-p_\text{flip})p_{l,0}^{r_t-r_l-1}}{p_{\text{tot.},r,r_l}}\\
   & \forall r_l<r_t< r, \nonumber\\
   & p_{\text{pot. loss},r_l,r_t,r,n_\text{CZ}^\text{stab.}+2}=\nonumber\\
    &\frac{p_{l,n_\text{CZ}^\text{stab.}+2}p_{\text{flip}}^{r-r_t-1}(1-p_{\text{flip}})p_{l,0}^{r_t-r_l-1}}{2p_{\text{tot.},r,r_l}}\\
  &  p_{\text{pot. loss},r_l,r,r,n_\text{CZ}^\text{stab.}+2}=\frac{p_{l,n_\text{CZ}^\text{stab.}+2} p_{l,0}^{r-r_l-1}}{2p_{\text{tot.},r,r_l}}
\end{flalign}
where the term $p_{\text{flip}}^{r-r_t}$ (resp. $p_{\text{flip}}^{r-r_t-1}$) represents the probability to not be detected over $r-r_t$ rounds if the loss occurred between the first and the $n_\text{CZ}^\text{stab.}+1$-th CZ gate (resp. at the $n_\text{CZ}^\text{stab.}+2$-th CZ gate) within a cycle. $p_{l,0}^{r-r_l-1}$ is the probability that the atom is not lost over $r_t-r_l-1$ rounds and the term $1-p_{\text{flip}}$ gives the probability that the LDU of the round $r$ detects the loss. If the round $r$ coincides with the last round of the entire correction protocol, this term should be removed because all the qubit are measured in the last cycle, hence no LDU is applied.

\section{Logical error probability in the $X$-basis}
\label{ap:xbasis}

For the sake of completeness, we present the logical error probability of both LDU protocols when the initial state is prepared in the $\ket{+}_L$ and the final state measured in the $X$ direction.

\begin{figure}[!!h]
    \centering
    \includegraphics[width=\linewidth]{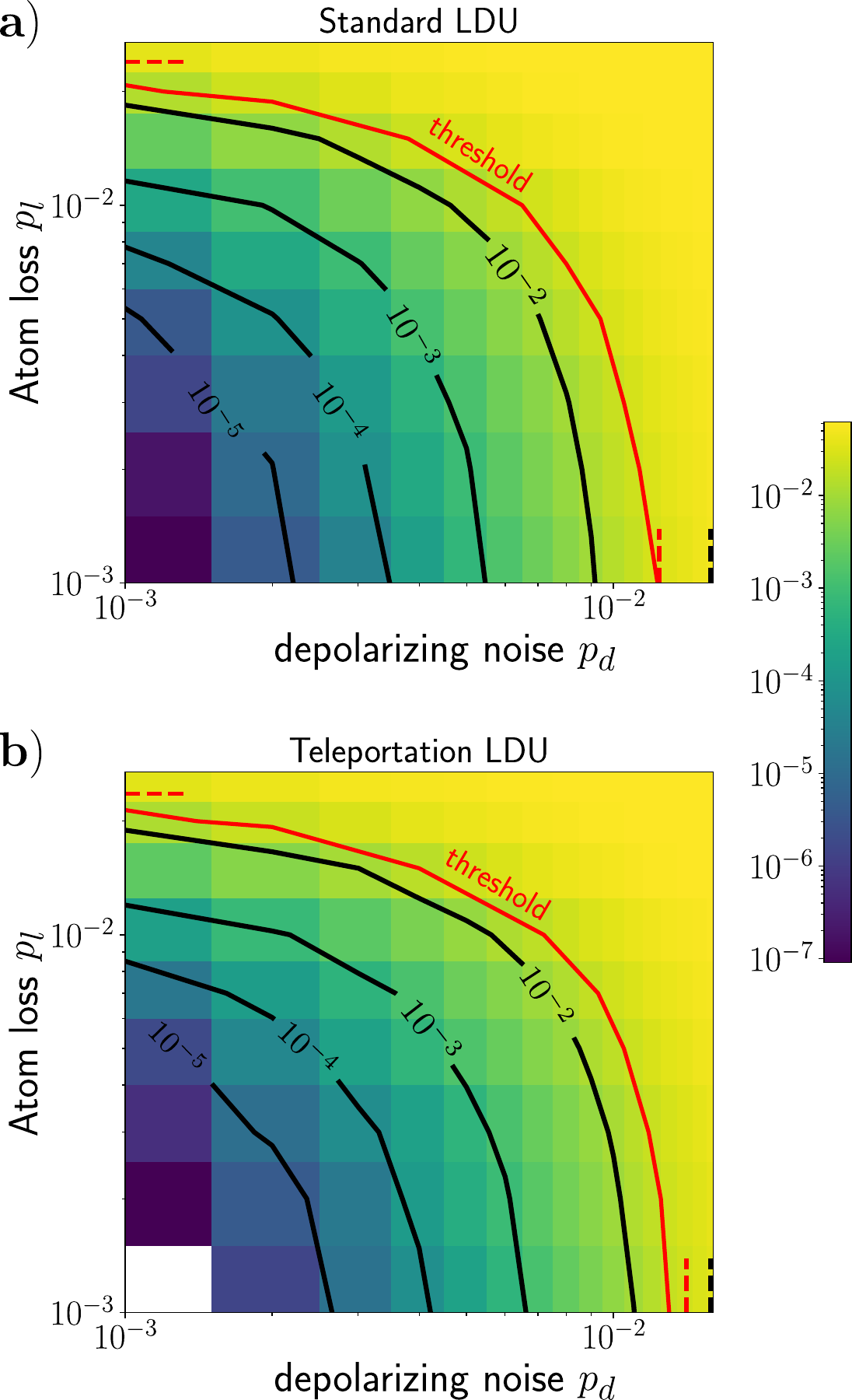}
    \caption{The logical error probability normalized by the number of rounds for a surface code of distance $d=11$ and $d$ cycles of stabilizer measurements as a function of the loss probability $p_l$ and the depolarizing noise error probability $p_d$. The initial state is prepared in $\ket{+}_L$. ${\bf a)}$ corresponds to the standard LDU protocol, and ${\bf b)}$ represents the teleportation-based LDU protocol. Each logical error probability was evaluated with $10^5$ shots, except for regions with $p_l<3. 10^{-3}$ and $p_d<2. 10^{-3}$, where $10^6$ shots were employed. The solid red line marks the error threshold while the solid black lines show curves of constant logical error probability at $10^{-2},10^{-3},10^{-4},10^{-5}$.  The horizontal and vertical dashed red lines indicate the threshold at vanishing depolarizing noise and loss probability, respectively. For comparison, the threshold of the standard surface code is shown by a black dashed line. In {$\bf b)$}, the blank region indicates no error was found.}
    \label{fig:xbasis}
\end{figure}

 Qualitatively bottom panels of Fig.~\ref{fig:error_loss_depo} and Fig.~\ref{fig:xbasis} show similar trends. The scaling derived for the initial state $\ket{0}_L$ also applies to $\ket{+}_L$, though there is a slight difference in the threshold. At vanishing depolarizing noise, the threshold for both protocols is around $2.4\%$, which is slightly lower than for the $\ket{0}_L$ state. This asymmetry between the $Z$-basis and $X$-basis may be attributed to the fact that the fresh atoms were initialized in the $\ket{0}$ state. 
 %Additionally, in the teleportation-based protocol, measuring the older atoms leads to $Z$-errors $50\%$ of the time, further contributing to this asymmetry.
\section{Logical error probability in the Z-basis with linear scale axes}
\label{ap:linear}
\par
\begin{figure}[!!h]
    \centering
    \includegraphics[width=\linewidth]{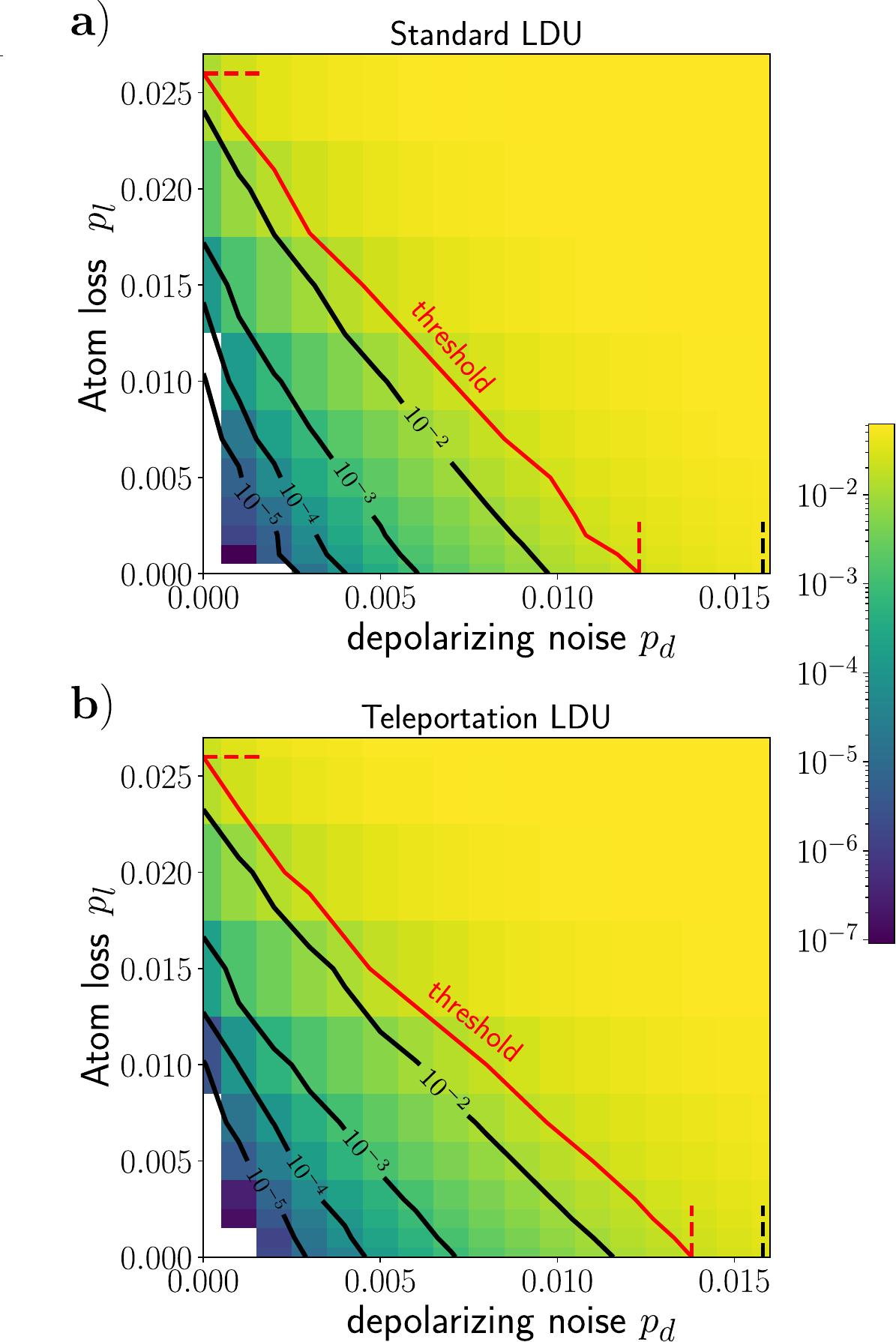}
    \caption{The logical error probability normalized by the number of rounds for a surface code of distance $d=11$ and $d$ cycles of stabilizer measurements as a function of the loss probability $p_l$ and the depolarizing noise error probability $p_d$. It is the same plots as in Fig.~\ref{fig:error_loss_depo} {$\bf c)$},{$\bf d)$} but with linear scale axes to point out the linear scaling of the iso-logical error probability curves. The blank regions indicate no errors were found.       }
    \label{fig:linear}
\end{figure}

 In this Appendix, we motivate the derived ansatz function of Eq.~\eqref{eq:fit2}, by plotting the logical error probability normalized by the number of rounds for a distance code $d=11$ and both LDU protocols as a function of the loss probability $p_l$ and the depolarizing noise error probability $p_d$ on linear scale axes to highlight that curve of iso-logical error probability have an almost linear dependence in $p_l$ and $p_d$ (see Fig.~\ref{fig:linear}).
\section{Error mechanism combining atom loss and depolarizing noise }
\label{ap:combined errors}

\begin{figure}[!!h]
    \centering
    \includegraphics[width=0.7
    \linewidth]{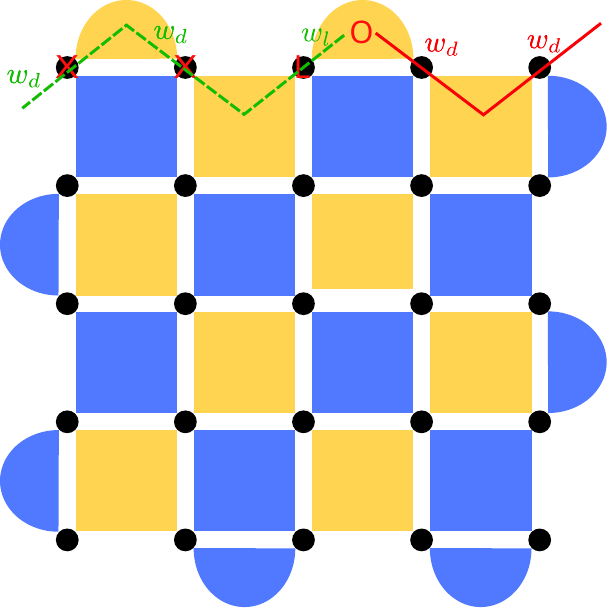}
    \caption{Example of a logical error on the surface code $d=5$. Black dots represent data qubits. Orange plaquettes indicate $Z$ stabilizers and blue plaquettes are $X$ stabilizers. Red X indicate an $X$ error on the corresponding qubit and red L marks the loss of the qubit. The red O represents the flipped stabilizer. The green dashed line shows the correct recovery procedure and the red solid line correspond to the incorrect one. $w_d$ and $w_l$ are the weights associated to the edges.}
    \label{fig:surface_code}
\end{figure}

We present an example of an error mechanism, for a surface code with $d=5$, (see Fig.~\ref{fig:surface_code}) which combine two $X$ errors and an atom loss indicated by red $X$ and $L$, respectively. We assume these errors arising at the beginning of the cycle. When considered independently, each error mechanism  introduces fewer than $\frac{d+1}{2}=3$ errors. However, atom loss and depolarizing noise combine such that the decoding procedure fails.
In this example, the three errors align such that only a single stabilizer flips, indicated by the red O. Using the MWPM algorithm, the path with the lowest weight $2w_d$ is selected (red solid line), where $w_d=\text{ln}\frac{1-p_d}{p_d}>0$. However, the correct recovery operator should follow the green dashed line with weight $2w_d+w_l$ where the weight $w_l=\text{ln}\frac{1-p_\text{loss}}{p_\text{loss}}$ with $p_\text{loss}=\frac{p_l}{2 p_\text{tot.}}(1-p_d)+p_d(1-\frac{p_l}{2 p_\text{tot.}})$ is computed using the \texttt{merge\_strategy=independent} option of Pymatching. In this case, we have $w_d>w_l>0$. Here $p_\text{tot.}$ depends on the LDU protocol employed and is the sum of the probabilities for all potential loss locations that trigger the corresponding LDU (see Appendix~\ref{ap:decoding} for details).
\section{Logical error for the naive decoder}
\label{ap:naive error}

\begin{figure}[!h]
\includegraphics[width=\linewidth]{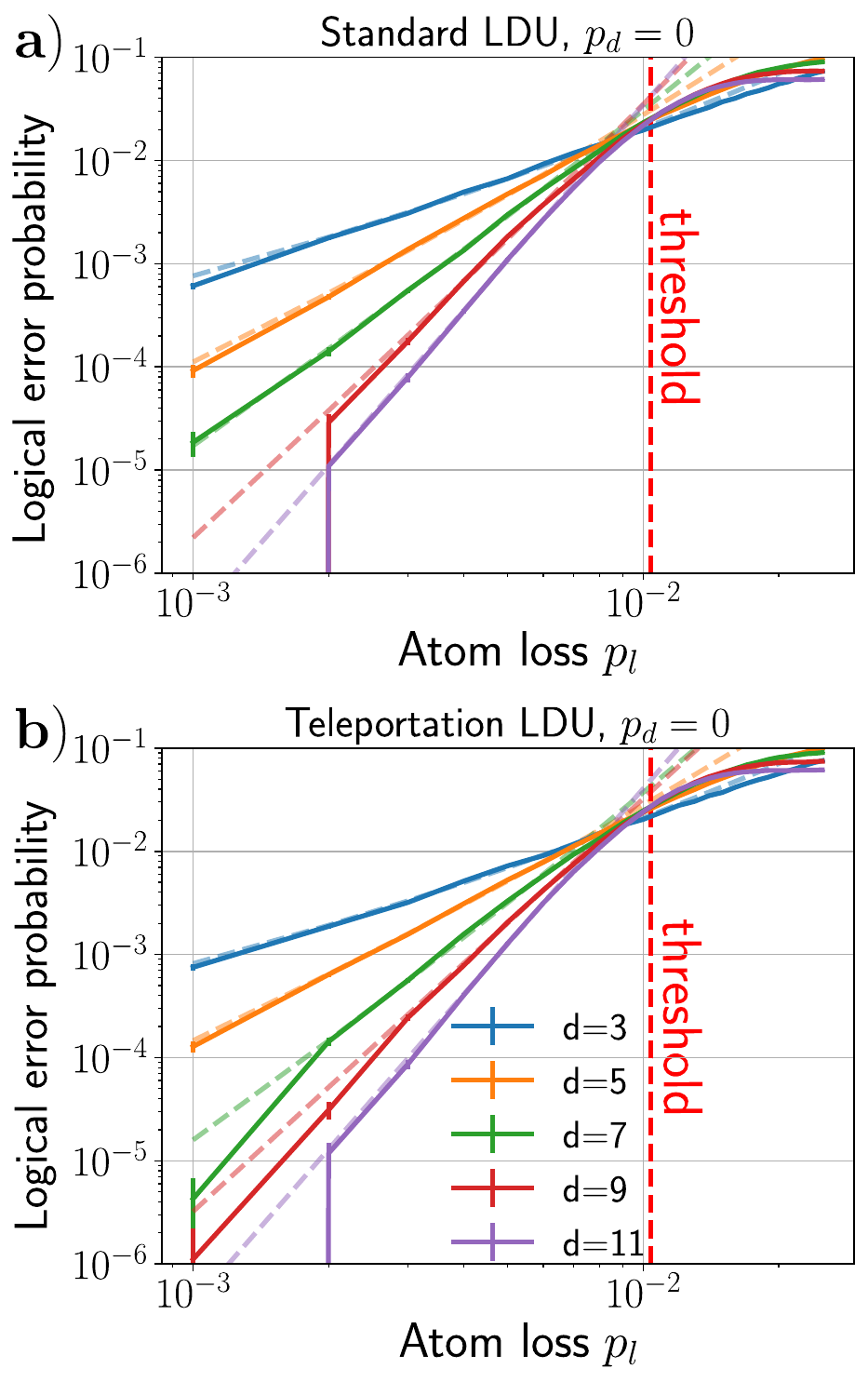} 

    \caption{Logical error probability normalized by the number of rounds employing the naive decoder at vanishing depolarizing noise as a function of the loss probability for code distance $d=3,5,7,9,11$ and $d$ cycles of stabilizer measurements for ${\bf a)}$ the standard LDU protocol and ${\bf b)}$ the teleportation LDU protocol. The dashed red vertical line indicates the threshold around $1\%$ for both protocols. Other colored dashed lines represent fits using Eq.~\eqref{eq:fit naive}. $10^5$ shots were used to estimate the logical error probability.}
    \label{fig:naive no depo}
\end{figure}

\begin{figure}[!!h]
    \centering
    \includegraphics[width=
    \linewidth]{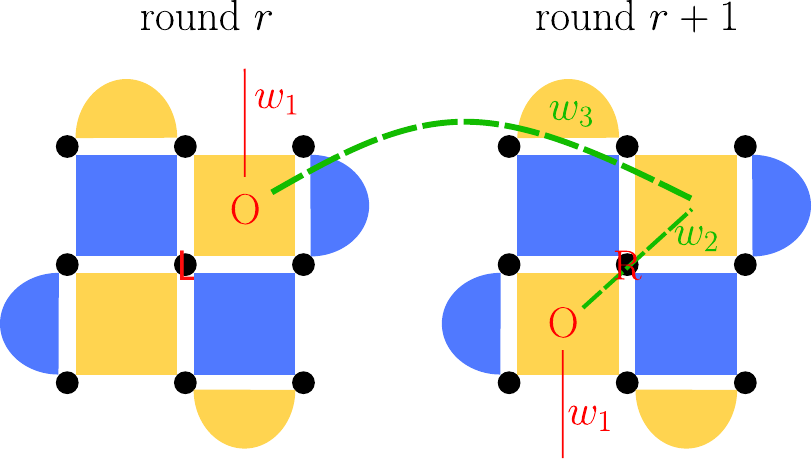}
    \caption{Example of a logical error employing the naive decoder on the surface code $d=3$. Black dots represent data qubits. Orange plaquettes indicate $Z$ stabilizers and blue plaquettes are $X$ stabilizers. The red L marks the loss of the qubit at round $r$ and the red R indicate the reloading at the subsequent cycle. The red O represents the flipped stabilizer. The green dashed line shows the correct recovery procedure and the red solid line correspond to the incorrect one. $w_1$, $w_2$ and $w_3$ are the weights associated to the edges. The width of the edges scales with the value of $w_i$: $w_1<w_2<w_3$.}
    \label{fig:naive error}
\end{figure}

In Fig.~\ref{fig:naive no depo} we plot the logical error probability normalized by the number of rounds at vanishing depolarizing noise and finite loss probability for both LDU protocols and using the naive decoder (see solid lines). The simulation is performed for surface codes of distances $d=3,5,7,9,11$ and $d$ cycles of stabilizer measurements. With the naive approach (see Sec.~\ref{sec:perf decoder}), losses are not treated as an erasure channel, leading us to expect a power law with an exponent of $\frac{d+1}{2}$; however, the observed scaling is even less favorable. We find that the smallest-weight physical errors leading to a logical error are of weight $\lceil\frac{d+1}{4}\rceil$ where $\lceil \cdot \rceil$ denotes the ceiling function. An example of such a logical error is illustrated in Fig.~\ref{fig:naive error} for a code distance $d=3$.

When a loss event occurs, the surrounding stabilizers may be flipped. In addition, when a new atom is reloaded in place of the lost one in the following cycle, it is, initially, in a product state with the rest of the code. Through stabilizer operations, it becomes entangled, but the measurement outcome of the neighbouring stabilizers may be flipped compared to the previous cycle. Thus, a combination of loss and reloading can introduce a weight-two error, reducing the effective logical error threshold to weight $\lceil\frac{d+1}{4}\rceil$ rather than the standard $\frac{d+1}{2}$.

In the specific scenario illustrated in Fig.~\ref{fig:naive error}, the middle atom is lost during the second CZ gate at round $r$ and subsequently reloaded at round $r+1$. The observed syndrome, indicated by the two red $O$ symbols, arises from the dashed green edges.  The loss and subsequent reloading trigger the flipping of the top-right orange Z stabilizer at both rounds $r$ and $r+1$. This simultaneous flipping of stabilizers at these cycles only occurs if the middle qubit loss happens precisely during the second CZ gate at round $r$, a scenario with low probability. In addition, at round $r+1$, both the top-right and bottom-left orange Z stabilizers might flip together, which can occur if the middle qubit is lost during any CZ gate at round $r$  or during the first CZ gate at round $r+1$. 

However, the observed syndrome is more likely to result from other loss mechanisms. Typically the flipped stabilizer at round $r$  can be caused by a loss of the upper right  qubit or the middle right qubit at round $r-1$ and during any of the CZ gates or at round $r$ but specifically  if the loss occurs during the first CZ gate. Likewise, the flipped stabilizer at round $r+1$ might stem from the loss of the middle-left or bottom-left qubit at round $r$ during any CZ gate, or at round $r+1$ if the loss occurs during the first CZ gate.
Overall, the observed syndrome is much more likely to be obtained by the red edges that the green ones causing a logical error.\\

However, these logical errors are rare, therefore higher order terms up to the weight $\frac{d+1}{2}$ must be included in the expansion of the logical error probability function as they cannot be neglected. Consequently, we fit the logical error probability in Fig.~\ref{fig:naive no depo} (see dashed lines) with a function of the form:
\begin{align}
    \sum_{k= \lceil \frac{d+1}{4}\rceil }^{\frac{d+1}{2}} b_{k}(d) p_l^k
    \label{eq:fit naive}
\end{align}
where $b_k(d)$  are fitting parameters which depends on the distance code. This functional form allows us to accurately capture the behavior of the logical error probability when using the naive decoder.
\end{document}